\documentclass[12pt,twoside]{article}

\pdfoutput=1

\usepackage[usenames]{color}
\usepackage{lscape}
\usepackage{amssymb}
\usepackage{rotating}
\usepackage[T1]{fontenc}
\usepackage{graphicx}
\usepackage{fancyhdr}
\usepackage{amsmath}
\usepackage{ae}
\usepackage{aecompl}
\usepackage{hyperref}
\def\pa{\partial}
\def\nn{\nonumber \\}

\def\hathh{\hat{h}\hat{h}}
\def\hatHH{\hat{H}\hat{H}}
\def\hatss{\hat{s}\hat{s}}
\def\hathH{\hat{h}\hat{H}}
\def\haths{\hat{h}\hat{s}}
\def\hatHs{\hat{H}\hat{s}}
\def\hataa{\hat{a}\hat{a}}
\def\hatAa{\hat{A}\hat{a}}
\def\hatAA{\hat{A}\hat{A}}

\def\mlsp{m_{\rm LSP}}

\newlength{\dinwidth}
\newlength{\dinmargin}
\begin{document}

\vspace*{1cm}

\begin{center}
\Large\bf Spin-dependent constraints on blind spots for thermal 
singlino-higgsino dark matter with(out) light singlets
\end{center}

\vspace*{2mm}

\vspace*{5mm} \noindent
\vskip 0.5cm
\centerline{\bf
Marcin Badziak${}^{a,b,}$\footnote[1]{mbadziak@fuw.edu.pl},
Marek Olechowski${}^{a,}$\footnote[2]{Marek.Olechowski@fuw.edu.pl},
Pawe\l\ Szczerbiak${}^{a,}$\footnote[3]{Pawel.Szczerbiak@fuw.edu.pl}
}
\vskip 5mm

\centerline{${}^a$\em Institute of Theoretical Physics,
Faculty of Physics, University of Warsaw} 
\centerline{\em ul.~Pasteura 5, PL--02--093 Warsaw, Poland} 
\centerline{${}^b$\em Berkeley Center for Theoretical Physics, Department of Physics,}
\centerline{\em and Theoretical Physics Group, Lawrence Berkeley National Laboratory,}
\centerline{\em University of California, Berkeley, CA 94720, USA}

\vskip 1cm

\centerline{\bf Abstract}
\vskip 3mm

The LUX experiment has recently set very strong constraints on spin-independent interactions of WIMP with nuclei. These null results can be
accommodated in NMSSM provided that the effective spin-independent coupling of the LSP to nucleons is suppressed. We investigate thermal relic
abundance of singlino-higgsino LSP in these so-called spin-independent blind spots and derive current constraints and prospects for direct detection
of spin-dependent interactions of the LSP with nuclei providing strong constraints on parameter space. We show that if the Higgs boson is the only
light
scalar the new LUX constraints set a lower bound on the LSP mass of about 300 GeV except for a small range around the half of $Z^0$ boson
masses
where resonant
annihilation via $Z^0$
exchange dominates. XENON1T will probe entire range of LSP masses except for a tiny $Z^0$-resonant region that may be
tested
by the LZ experiment.
These conclusions apply to general singlet-doublet dark matter annihilating dominantly to $t\bar{t}$.
Presence of light singlet (pseudo)scalars generically relaxes the constraints because new LSP (resonant and non-resonant) annihilation channels
become important. 
Even away from resonant regions, the lower limit on the LSP mass from LUX is relaxed to about 250 GeV while XENON1T may not be sensitive to the LSP
masses above about 400 GeV.

\newpage

\section{Introduction}

Weakly Interacting Massive Particle (WIMP) has been considered as one of 
the most attractive candidates for dark matter (DM). WIMPs have been
intensively searched for in direct detection experiments and the limits 
on the DM scattering cross-section on nuclei improved by several orders of
magnitude over the last decade. Currently the most constraining limits 
come from the LUX \cite{LUX2016} experiment which will be soon improved by the
XENON1T \cite{Xenon1T} experiment which is expected to be superseded 
by the LZ \cite{LZ} experiment in near future. In many theories, WIMP 
interactions with nuclei are mediated by the Higgs and $Z^0$ bosons. 
The electroweak-strength couplings of WIMP to the Higgs and $Z^0$ bosons 
were excluded already few years ago by XENON100 \cite{Xenon100} as noted 
in Ref.~\cite{bs_Hall}. The recent LUX constraints pushed models to regions 
of parameter spaces where these couplings are strongly suppressed i.e.~to 
the vicinity of so-called blind spots in direct detection.

Supersymmetric (SUSY) extensions of the Standard Model (SM) generically 
provide a WIMP candidate in the form of the lightest neutralino which is 
often the lightest sparticle (LSP). Some earlier studies of neutralino DM 
include Refs.~\cite{Ellwanger:2014dfa}-\cite{well-tempered}.
 Blind spots for neutralino DM has been identified e.g.~in 
Refs.~\cite{bs_Hall,bs_Wagner,bs_Crivellin,Han:2016qtc} for 
Minimal Supersymmetric Standard Model (MSSM)
and in Refs.~\cite{BS_NMSSM,Cheung:2014lqa} for 
Next-to-Minimal Supersymmetric Standard Model (NMSSM). 
Several recent papers emphasized a big impact of new LUX constraints on 
the spin-independent (SI) scattering cross-section on the parameter space
 of MSSM \cite{welltemperedBOS,Huang:2017kdh} and NMSSM
\cite{Cao:2016cnv,Ellwanger:2016sur,Beskidt:2017xsd}. A universal conclusion 
of these papers is that viable points in the parameter space still exist but
they reside very close to blind spots for SI scattering cross-section.

In the present article we study implications of the assumption that the 
SI scattering of the LSP is so small (below the neutrino background) 
that probably it will never be detected in direct detection of its SI 
interactions with nuclei. We also demand that the LSP has thermal relic 
abundance in agreement with the Planck measurement
$\Omega h^2\approx0.12$ \cite{Planck}. These two assumptions lead to 
interesting predictions for the model parameter space. We focus on NMSSM with
singlino-higgsino LSP but many of our conclusions are valid also for more 
general singlet-doublet DM models (studied e.g.~in
Refs.~\cite{Cohen:2011ec}-\cite{Banerjee:2016hsk}). We investigate how the resulting parameter space can be constrained by direct detection
experiments focusing on the spin-dependent (SD) LSP interactions with nuclei. We assess the impact of new LUX results presented at the Moriond
2017 conference \cite{LUX_SD_n} on the parameter space, as well as sensitivity of future XENON1T and LZ experiments. Rather than doing huge
numerical scans of the NMSSM parameter space we study separately several classes of SI blind spots identified in Ref.~\cite{BS_NMSSM}. This way it
becomes possible to understand which effects have the biggest impact on the constraints. In particular, we emphasize the role of light singlets in
relaxing the constraints.

The rest of the article is organized as follows. In section~\ref{sec:model} we introduce the model and conventions used. In the next two sections we
focus on general NMSSM.  In section~\ref{sec:bs_fh} we discuss the case in which singlet-like states are heavy and SI cross-section is solely
determined by the exchange of the Higgs boson. In section~\ref{sec:bs_fhs} we discuss how both the relic density and blind spot condition is affected
by the presence of light singlets. In section~\ref{sec:Z3} we analyze the 
$\mathbb{Z}_3$ invariant NMSSM. We reserve section~\ref{sec:concl} for our
conclusions.

\section{Model and conventions}
\label{sec:model}

In this section we collect formulae useful for the analysis 
performed in the rest of this work. We adopt conventions 
as in the previous paper \cite{BS_NMSSM} 
where more details may be found.
We start with the most general NMSSM with the superpotential 
and the soft terms given by
\begin{align}
\label{W}
W &= W^{\rm MSSM}+\lambda SH_uH_d + f(S) \,,
\\
-{\cal{L}}_{\rm soft}
&= -{\cal{L}}_{\rm soft}^{\rm MSSM}
+m_{H_u}^2\left|H_u\right|^2+m_{H_d}^2\left|H_d\right|^2+m_{S}^2\left|S\right|^2
\nn
&\label{Lsoft}
\quad+\left(
A_\lambda\lambda H_u H_d S +\frac13A_\kappa\kappa S^3
+m_3^2H_uH_d + \frac12 m_S'^2S^2 + \xi_SS
+{\rm h.c.}\right),
\end{align}
where $S$ is a chiral SM-singlet superfield. 
In the simplest version, known as the scale-invariant 
or $\mathbb Z_3$-symmetric NMSSM, 
$m_3^2=m_S'^2=\xi_S=0$ while $f(S)\equiv\kappa S^3/3$. In more general 
models $f(S)\equiv\xi_F S+\mu'S^2/2+\kappa S^3/3$.

The mass squared matrix for the neutral CP-even scalar fields, 
in the basis $\left(\hat{h}, \hat{H}, \hat{s}\right)$ 
related to the interaction basis by a rotation by angle $\beta$ 
(see \cite{BS_NMSSM} for details), reads:
\begin{equation}
\label{M_s^2}
 {M}_s^2=
\left(
\begin{array}{ccc}
  {M}^2_{\hathh} & {M}^2_{\hathH} & {M}^2_{\haths} \\[4pt]
   {M}^2_{\hathH} & {M}^2_{\hatHH} & {M}^2_{\hatHs} \\[4pt]
   {M}^2_{\haths} & {M}^2_{\hatHs} & {M}^2_{\hatss} \\
\end{array}
\right) \,,
\end{equation}
where
\begin{align}
\label{Mhh}
 &{M}^2_{\hathh} = M_Z^2\cos^2\left(2\beta\right)
+ \lambda^2 v_h^2\sin^2\left(2\beta\right) 
+\Delta_{\hathh}, \\
\label{MHH}
&{M}^2_{\hatHH} = (M_Z^2-\lambda^2 v_h^2)\sin^2\left(2\beta\right) 
+ 
\frac{2}{\sin\left(2\beta\right)}
\left(\mu A_{\lambda} + 
\frac{\mu\langle\pa_S f\rangle}{v_s}
+m_3^2\right)+\Delta_{\hatHH}, 
\\
\label{Mss}
&{M}^2_{\hatss} =  
\frac12 \lambda v_h^2 \sin2\beta
\left(\frac{A_{\lambda}+\langle\pa_S^2 f\rangle}{v_s}-\left<\partial^3_Sf\right>\right)
+\langle(\partial_S^2f)^2 + \partial_S f\,\partial_S^3 f\rangle
-\frac{\langle\partial_S f\,\partial_S^2 f\rangle}{v_s}
\nonumber\\ &\hspace{32pt}
+A_\kappa\kappa v_s -\frac{\xi_S}{v_s}
+\Delta_{\hatss}\,, \\
\label{MhH}
& {M}^2_{\hathH} = \frac{1}{2}(M^2_Z-\lambda^2 v_h^2)\sin4\beta
+\Delta_{\hathH}, \\
\label{Mhs}
& {M}^2_{\haths} =  
\lambda v_h (2\mu- \left(A_{\lambda}+\langle\pa_S^2 f\rangle\right) \sin2\beta)
+\Delta_{\haths}, \\
\label{MHs}
& {M}^2_{\hatHs} =
 \lambda v_h \left(A_{\lambda}+\langle\pa_S^2 f\rangle\right) \cos2\beta
+\Delta_{\hatHs} ,
\end{align}
$\Delta_{{\hat{h}_i}\hat{h}_j}$ are radiative corrections,
$v_s$, $v_h\sin\beta$ and $v_h\cos\beta$ are VEVs of the singlet 
and the two doublets, respectively.
The mass eigenstates of ${M}^2$ are
denoted by $h_i$ with $h_i=h,H,s$
($h$ is the 125 GeV scalar discovered by the LHC 
experiments). These mass eigenstates 
are expressed in terms of the hatted fields with the help of the 
diagonalization matrix $\tilde{S}$:
\begin{equation}
\label{tildeS}
h_i
=\tilde{S}_{h_i\hat{h}}\hat{h}
+\tilde{S}_{h_i\hat{H}}\hat{H}
+\tilde{S}_{h_i\hat{s}}\hat{s}
\,.
\end{equation}

The mass squared matrix for the neutral pseudoscalars, after rotating 
away the Goldstone boson, has the form
\begin{equation}
\label{M_p^2}
 {M}_p^2=
\left(
\begin{array}{cc}
   {M}^2_{\hatAA} & {M}^2_{\hatAa} \\[4pt]
   {M}^2_{\hatAa} & {M}^2_{\hataa} \\
\end{array}
\right) \,,
\end{equation}
where
\begin{align}
\label{MAA}
&{M}^2_{\hatAA} 
=
\frac{2}{\sin\left(2\beta\right)}
\left(\mu A_{\lambda} + 
\frac{\mu\langle\pa_S f\rangle}{v_s}
+m_3^2\right)+\Delta_{\hatAA}
\,, \\
\label{MAa}
&{M}^2_{\hatAa} 
= 
{\lambda v_h}  \left(A_{\lambda} - \langle\pa_S^2 f\rangle\right) 
+ \Delta_{\hatAa}
\,, \\
\label{Maa}
&{M}^2_{\hataa} 
= 
\frac12 \lambda v_h^2 \sin2\beta
\left(\frac{A_{\lambda}+\langle\pa_S^2 f\rangle}{v_s}
+\left<\partial^3_Sf\right>\right)
+\langle(\partial_S^2f)^2 - \partial_S f\,\partial_S^3 f\rangle
-\frac{\langle\partial_S f\,\partial_S^2 f\rangle}{v_s}
\nonumber\\ &\hspace{32pt}
-3A_\kappa\kappa v_s - 2m_S'^2-\frac{\xi_S}{v_s}
+ \Delta_{\hataa},
\,. 
\end{align}
Diagonalizing $M_p^2$ with the matrix $\tilde{P}$ 
one gets the mass eigenvalues
$a_j=a,A$:
\begin{equation}
\label{tildeP}
a_j=\tilde{P}_{{a_j}\hat{a}}\hat{a}+\tilde{P}_{{a_j}\hat{A}}\hat{A}
\,.
\end{equation}

After decoupling of the gauginos (assumed in this work) 
the neutralino mass sub-matrix describing the three lightest
states takes the form:
\begin{equation}
\label{M_chi}
 {M_{\chi^0}}=
\left(
\begin{array}{ccc}
  0 & -\mu & -\lambda v_h \sin\beta \\[4pt]
  -\mu & 0 & -\lambda v_h \cos\beta \\[4pt]
  -\lambda v_h \sin\beta & -\lambda v_h \cos\beta 
& \langle\partial_S^2f\rangle \\
\end{array}
\right) \,.
\end{equation}
Trading the model dependent term $\langle\partial_S^2f\rangle$
for one of the eigenvalues, $m_{\chi_j}$, of the above neutralino mass
matrix we find the following (exact at the tree level) relations 
for the neutralino diagonalization matrix 
elements:
\begin{align}
\label{Nj3Nj5}
\frac{N_{j3}}{N_{j5}}
=
\frac{\lambda v_h}{\mu}
\,
\frac{(m_{\chi_j}/\mu)\sin\beta-\cos\beta}
{1-\left(m_{\chi_j}/\mu\right)^2}
\,,\\[4pt]
\label{Nj4Nj5}
\frac{N_{j4}}{N_{j5}}
=
\frac{\lambda v_h}{\mu}
\,
\frac{(m_{\chi_j}/\mu)\cos\beta-\sin\beta}
{1-\left(m_{\chi_j}/\mu\right)^2}
\,,
\end{align}
where $N_{j3}$, $N_{j4}$ and $N_{j5}$ denote, respectively, the two higgsino 
and the singlino components of the $j$-th neutralino mass eigenstate
while $j=1,2,3$ and $|m_{\chi_1}|\le|m_{\chi_2}|\le|m_{\chi_3}|$.
Using the last two equations and neglecting contributions from decoupled 
gauginos one can express the composition of the three lighter neutralinos 
in terms of: $\lambda v_h$, $m_\chi/\mu$ and $\tan\beta$.
Later we will be interested mainly in the LSP corresponding to $j=1$, 
so to simplify the notation we will use $m_{\chi}\equiv m_{\chi_1}$. 
The physical (positive) LSP mass is given by $m_{\rm LSP}\equiv|m_{\chi}|$.

In the present work we are interested mainly in two properties
of the LSP particles: their cross-sections on nucleons and 
their relic abundance. 
The spin-dependent LSP-nucleon scattering cross section 
is dominated by the $Z^0$ boson exchange and equals
\begin{equation}
\label{eq:sigSD}
\sigma_{\rm SD}^{(N)}=C^{(N)}\cdot 10^{-38}\;{\rm cm^2}\;\,(N_{13}^2-N_{14}^2)^2\,,
\end{equation}
where $C^{(p)}\approx 4$, $C^{(n)}\approx 3.1$~\cite{SD}.
The spin-independent cross-section for the LSP interacting 
with the nucleus with the atomic number $Z^0$ and the mass number $A$ 
is given by
\begin{equation}
\sigma_{\rm SI}
=
\frac{4\mu^2_{\rm red}}{\pi}\,\frac{\left[Zf^{(p)}+(A-Z)f^{(n)}\right]^2}{A^2}
\,,
\end{equation}
where $\mu^2_{\rm red}$ is the reduced mass of the nucleus and the LSP. 
When the squarks are heavy, as assumed in the present work, 
the effective couplings $f^{(N)}$ ($N=p,n$) 
are dominated by the t-channel exchange of the CP-even scalars
\cite{Jungman:1995df}:
\begin{equation}
\label{fN}
f^{(N)}
\approx
\sum_{i=1}^3
f^{(N)}_{h_i}
\equiv
\sum_{i=1}^3
\frac{\alpha_{h_i\chi\chi}\alpha_{h_iNN}}{2m_{h_i}^2}
\,.
\end{equation}
Further details may be found in Appendix \ref{App:cross-sections}.
Formulae for the LSP annihilation cross section and its relic density 
are much more complicated (some of them are collected in Appendices 
\ref{App:annihilation} and \ref{App:relic_density}).

The main goal of the present work is to identify regions of the NMSSM 
parameter space for which the singlino-higgsino LSP particles fulfill 
three conditions: 
1) have very small, below the neutrino background, SI cross-section 
on nuclei  (SI blind spots);
2) have small SD cross-sections to be consistent with present 
experimental bounds; 
3) have relic density close to the experimentally favored 
value $\Omega h^2\approx0.12$ so the LSP can play the role of a dominant 
component of
DM.\footnote{In order to take into account theoretical uncertainties in the 
relic abundance calculations we consider parameters leading to 
$\Omega h^2 = 0.12 \pm 0.02$ when calculated with the help of 
{\tt MicrOMEGAs 4.3.1} \cite{micromegas} and 
{\tt NMSSMTools 5.0.2}~\cite{NTools1,NTools2} which we use to calculate the NMSSM spectrum.}
 Of course, such points in the parameter space
must be consistent with other experimental constrains 
e.g.~those derived from the LHC \cite{Higgscomb} and LEP data \cite{LEPchargino,LEPZinv}. In the next sections we 
discuss solutions fulfilling all the above mentioned 
conditions, starting with the simplest case of blind spots without 
interference effects and then for blind spots for which such 
effects are crucial. We investigate also modifications present 
in $\mathbb{Z}_3$ invariant NMSSM.

\section{Blind spots without interference effects and relic density 
\label{sec:bs_fh}}

In this section we consider situation when the SI blind spot (BS)
takes place without interference effects i.e.~when all three 
contributions to the effective coupling $f^{(N)}$ in eq.~\eqref{fN} 
are very small. Two of them are small because the corresponding 
scalars, $s$ and $H$, are very heavy while $f_{h}$ is suppressed 
due to smallness of $\alpha_{h\chi\chi}$.
In the next subsection we discuss the simplest case in which the 
mixing among the scalars may be neglected. Then the effects of 
such mixing will be taken into account.

\subsection{Without scalar mixing \label{ssec:bs_fh_nomix}}

\begin{figure}
\center
\includegraphics[width=0.8\textwidth]{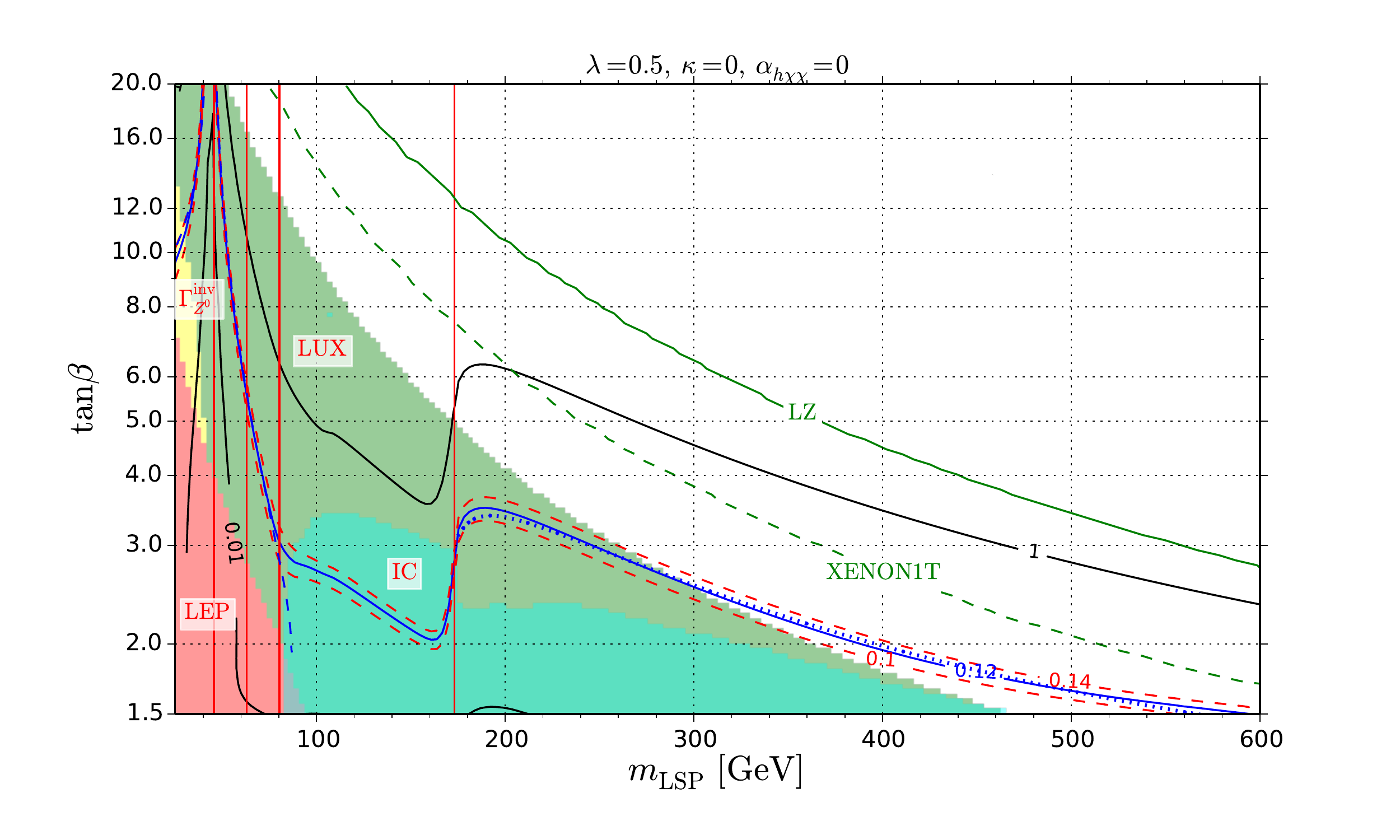}
\caption{Contour lines of $\Omega h^2$ (obtained with 
{\tt MicrOMEGAs 4.3.1} \cite{micromegas} and 
{\tt NMSSMTools 5.0.2}~\cite{NTools1,NTools2})
as functions of $m_{\rm LSP}$ and $\tan\beta$ for blind spots 
with decoupled all Higgs particles except the SM-like one. 
Dashed and dotted blue lines correspond to~eq.~\eqref{eq:res_Z_omega} 
and \eqref{eq:ann_tt_omega}, respectively, after 
substituting~\eqref{N13sqmN14sq_0}, for $\Omega h^2=0.12$. 
Red region depicts the points with $m_{\chi_1^\pm}<103$~GeV which 
are ruled out by LEP \cite{LEPchargino}. Yellow area is forbidden because of
eq.~\eqref{eq:Z_ogr},
whereas green/cyan one due to LUX/IceCube (IC) limits on spin-dependent LSP interaction with nucleons. Vertical red lines correspond to (from left to right)
$m_{\rm LSP}=m_{Z^0}/2,\,m_h/2,\,m_W,\,m_t$.
}
\label{fig:cont_om_0}
\end{figure}

When the interference effects and scalar mixing may be neglected 
the SI blind spot condition has the following simple form:
\begin{equation}
\label{bs_h_0}
\frac{m_\chi}{\mu}-\sin2\beta=0\,,
\end{equation}
which corresponds to vanishing Higgs-LSP-LSP coupling for $m_s,m_H\to\infty$. 
In Fig.~\ref{fig:cont_om_0} the dependence of $\Omega h^2$ on $m_{\rm LSP}$ 
and $\tan\beta$ for such blind spots is shown for some specific values of 
$\lambda$ and $\kappa$ (the latter parameter does not influence the situation 
as long as the resonance with the lightest pseudoscalar is not considered).
Some parts of the $(m_{\rm LSP},\tan\beta)$ plane are excluded by the upper  
bounds on $\sigma_{\rm SD}$ from LUX \cite{LUX_SD_n} and IceCube 
\cite{IceCubeNEW} experiments (and also by the LEP data). 
Particularly important are the new LUX constraints which exclude 
large part of the parameter space with $\Omega h^2=0.12\pm0.02$.
We note that in the allowed part of the  parameter space presented 
in Fig.~\ref{fig:cont_om_0} correct thermal relic abundance is obtained for
singlino-dominated LSP.
It is not difficult to understand the 
results shown in Fig.~\ref{fig:cont_om_0} using (approximate) 
analytic formulae.

For given values of $m_\chi$ and $\tan\beta$, the blind spot condition 
\eqref{bs_h_0} together with eqs.~\eqref{Nj3Nj5} and \eqref{Nj4Nj5} 
may be used to obtain the LSP composition. 
For example, the combination 
which determines the LSP coupling to the $Z^0$ boson is given 
by (see Appendix \ref{App:cross-sections})\footnote{
There are some corrections caused by not totally decoupled particles. 
In our numerical scan we took $M_1,\,M_2\approx 4$~TeV, 
$m_H,\,m_s\approx 4$~TeV, $m_{\tilde{q}}\approx 3$~TeV.}
\begin{equation}
\label{N13sqmN14sq_0}
\left|N_{13}^2-N_{14}^2\right|=\left(\frac{\lambda v_h}{m_\chi}\right)^2
\frac{\sin^22\beta}{|\cos2\beta|}
\left[1+\left(\frac{\lambda v_h}{m_\chi}\right)^2\tan^22\beta\right]^{-1}\,.
\end{equation}
All three components (gauginos are decoupled) may be expressed in terms
of the above combination using
\begin{align}
&N_{13}^2=\cos^2\beta\,\frac{\left|N_{13}^2-N_{14}^2\right|}{\cos2\beta}
\,,\label{N13_0}
\\
&N_{14}^2=\sin^2\beta\,\frac{\left|N_{13}^2-N_{14}^2\right|}{\cos2\beta}
\,,\label{N14_0}
\\
&N_{15}^2=1-\frac{\left|N_{13}^2-N_{14}^2\right|}{\cos2\beta}
\,.\label{N15_0}
\end{align}
The above expressions are valid as long as blind spot condition 
\eqref{bs_h_0} is satisfied. 
For the LSP masses for which the annihilation cross section is 
dominated by the s-channel $Z^0$ exchange this is enough to calculate 
the LSP relic density to a good accuracy. The approximate formulae are
(see the Appendix \ref{App:annihilation} for the details): 
\begin{equation}
\label{eq:res_Z_omega}
\Omega h^2\approx 0.1\left(\frac{0.3}{N_{13}^2-N_{14}^2}\right)^2
\frac{m_{Z^0}^2}{4m_{\chi}^2}
\left[\left(\frac{4m_{\chi}^2}{m_{Z^0}^2}-1 +\frac{\bar{v}^2}{4}\right)^2+
\frac{\Gamma_Z^2}{m_{Z^0}^2}\right]\,,
\end{equation}
for $m_{\rm LSP}$ of order $m_{Z^0}/2$ (and below the $W^+W^-$ threshold)
and
\begin{equation}
\label{eq:ann_tt_omega}
\Omega h^2\approx 0.1\left(\frac{0.05}{N_{13}^2-N_{14}^2}\right)^2
\left[
\sqrt{1-\frac{m_t^2}{m_{\chi}^2}}+
\frac{3}{4}\frac{1}{x_f}
\left(1-\frac{m_t^2}{2m_{\chi}^2}\right)
\frac{1}{\sqrt{1-\frac{m_t^2}{m_{\chi}^2}}}
\right]^{-1}\,,
\end{equation}
above the $t\bar{t}$ threshold. They reproduce very well the relic density 
calculated numerically using \texttt{MicrOMEGAs}, as may be seen in 
Fig.~\ref{fig:cont_om_0}.

\begin{figure}
\center
\includegraphics[width=0.49\textwidth]{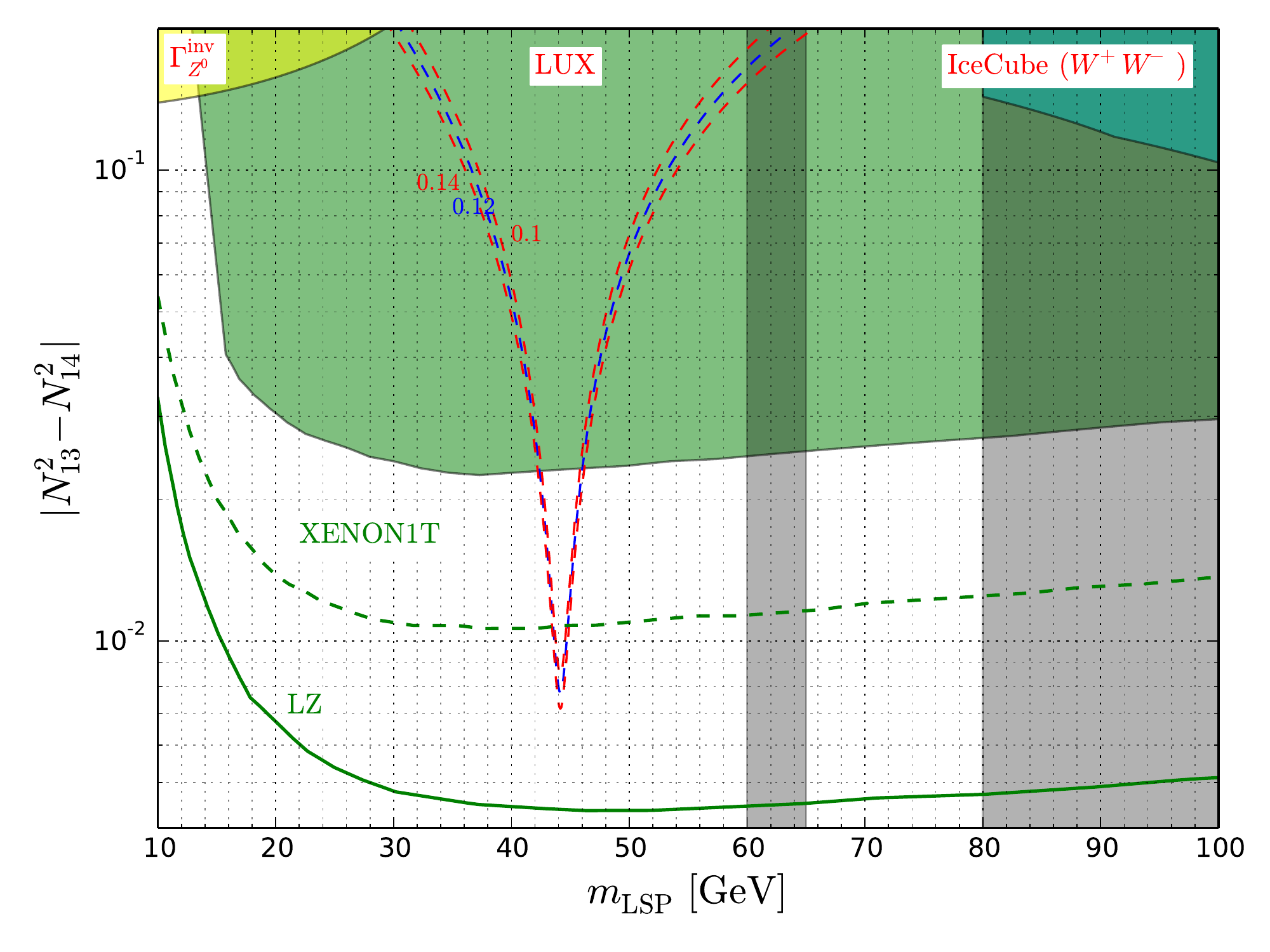}
\includegraphics[width=0.49\textwidth]{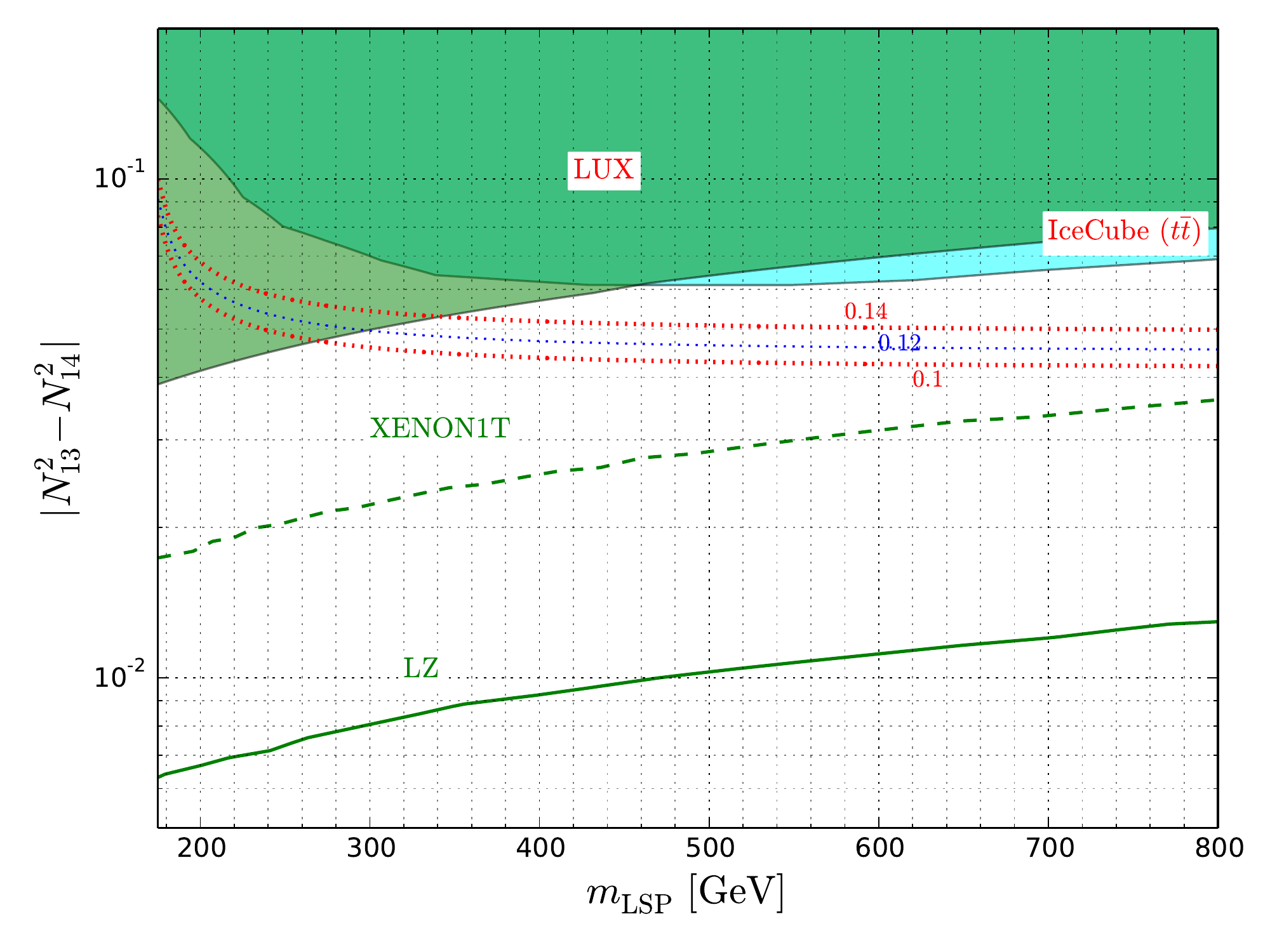}
\caption{Contour lines of $\Omega h^2=0.10,\,0.12,\,0.14$ as functions 
of $m_{\rm LSP}$ and $|N_{13}^2-N_{14}^2|$ for two LSP mass regimes 
(blue and red dashed lines in the left plot were obtained from 
eq.~\eqref{eq:res_Z_omega} whereas dotted lines in the right plot 
from eq.~\eqref{eq:ann_tt_omega}). Yellow area is
forbidden because of eq.~\eqref{eq:Z_ogr}. Grey color in the left 
plot denotes the regions in which $h$ resonance and $W^+W^-/Z^0Z^0$ 
channels may be important and affect the results. Green/cyan areas 
correspond to LUX/IceCube limits on spin-dependent LSP interaction with
nucleons~\cite{LUX_SD_n}, \cite{IceCubeNEW}. 
Dashed (continuous) green lines in both plots correspond 
to the precision of the future XENON1T (LZ) experiment of SD direct 
interaction of the LSP with neutrons~\cite{SD_future}.
}
\label{fig:res_Z_limits}
\end{figure}

The composition of the LSP is crucial not only for its relic density but 
also for some of the experimental constraints. One gets the following 
upper bounds on the combination \eqref{N13sqmN14sq_0}:
\begin{equation}
\label{eq:sigmaSD_ogr}
\left(N_{13}^2-N_{14}^2\right)^2
\lesssim
1.8\cdot10^{-3} \left(1-\frac{4m_{\chi}^2}{m_{Z^0}^2}\right)^{-3/2}
\end{equation}
from the LEP bound for invisible decays of $Z^0$~\cite{LEPZinv,Z_inv} and
\begin{equation}
\label{eq:Z_ogr}
\left(N_{13}^2-N_{14}^2\right)^2
\lesssim
\frac{\left(\sigma_{\rm SD}^{(N)}\right)^{\rm exp}}{C^{(N)}\cdot10^{-38}{\rm cm}^2}
\end{equation}
from experiments sensitive to the SD interactions of the LSP 
with protons or neutrons for which $C^{(p)}\approx 4$, $C^{(n)}\approx 3.1$, 
respectively~\cite{SD}. These upper bounds on $\left(N_{13}^2-N_{14}^2\right)^2$ 
as functions of the LSP mass are shown in Fig.~\ref{fig:res_Z_limits}. 

\begin{figure}[t]
\center
\includegraphics[width=0.49\textwidth]{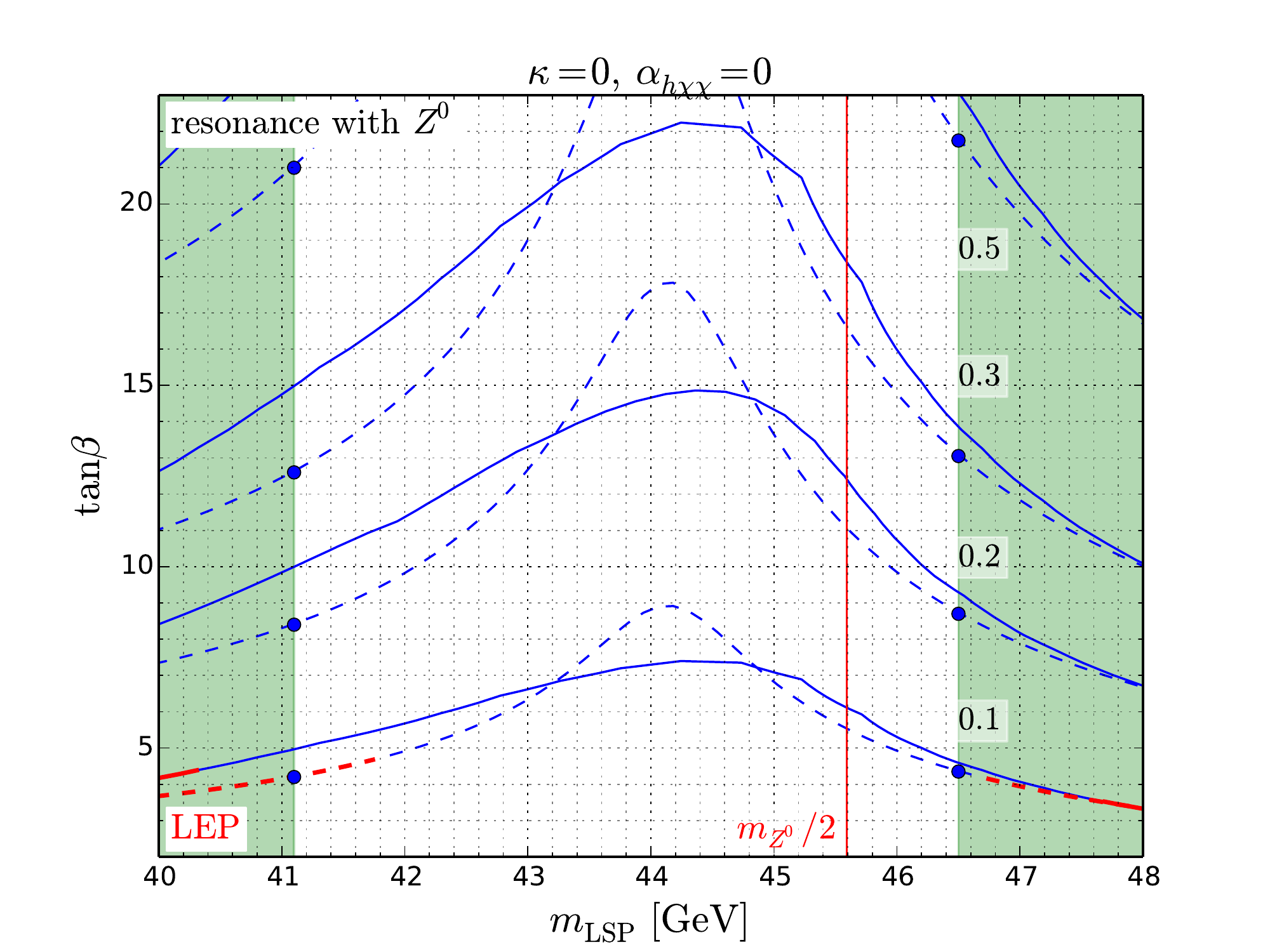}
\includegraphics[width=0.49\textwidth]{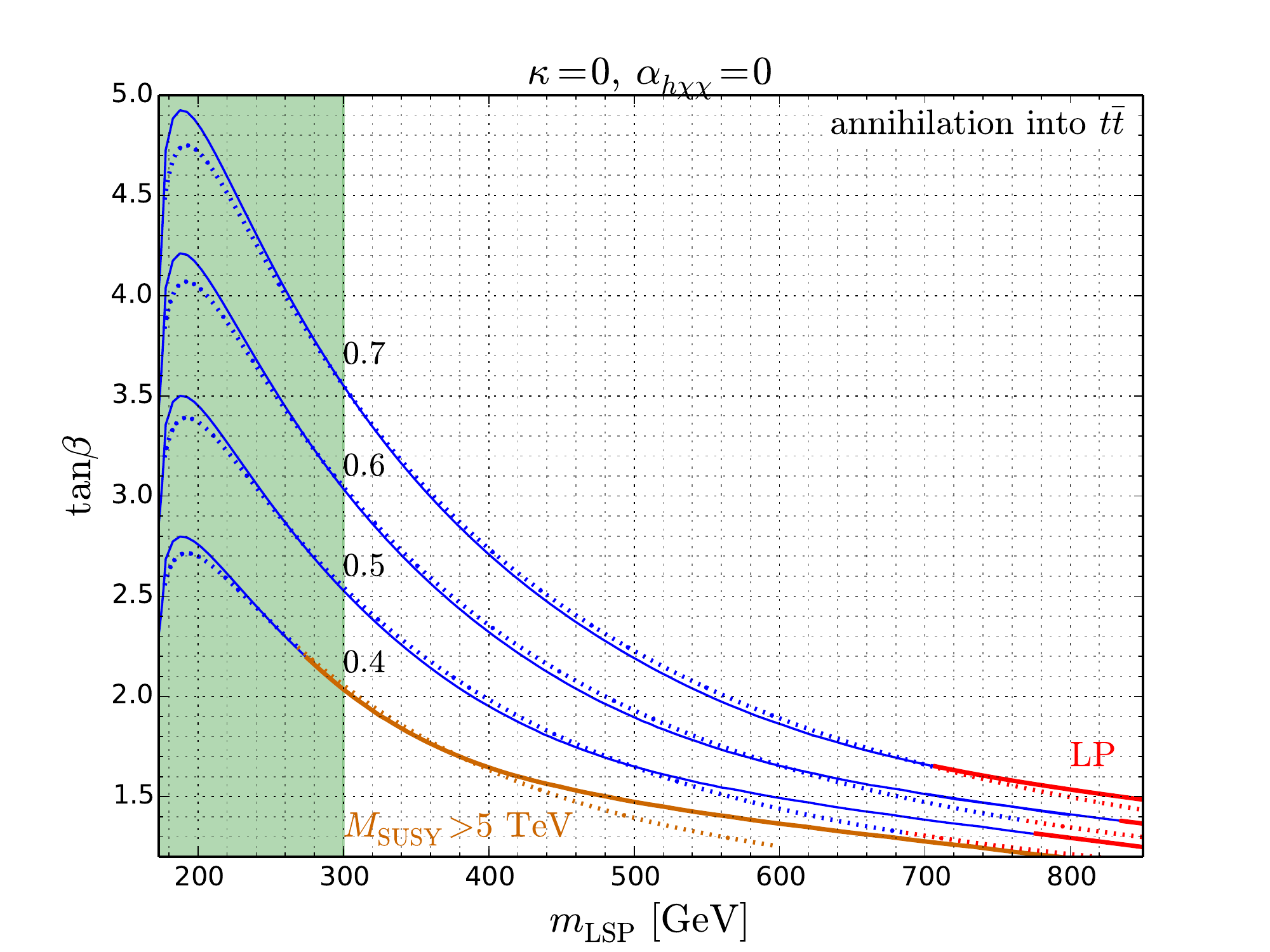}
\caption{Contour lines of $\Omega h^2=0.12$ (obtained with \texttt{MicrOMEGAs}) 
as functions of $m_{\rm LSP}$ and $\tan\beta$ for a few values of $\lambda$ 
(dashed lines in the left plot was obtained from eq.~\eqref{eq:res_Z_omega} 
after substituting~\eqref{N13sqmN14sq_0} whereas dotted lines in the right plot 
from eq.~\eqref{eq:ann_tt_omega} after substituting~\eqref{N13sqmN14sq_0}) 
for blind spots with decoupled all Higgs particles except the SM-like one. 
Left plot: green areas and thick red lines denote the points excluded by LUX 
and the LEP chargino searches, respectively;
thick blue points correspond to eq.~\ref{tanb_lam_1} and \ref{tanb_lam_2}. 
Right plot: green region and thick orange line depict the points excluded 
by LUX and points for which the stop masses above 5~TeV are necessary 
to obtain the correct Higgs mass (even when the contribution from the stop mixing is maximized), respectively. The red lines denoted 
by LP depict regions with a Landau pole below the GUT scale.
}
\label{fig:channel_Z_om_0}
\end{figure}

One can find the allowed range of $m_{\rm LSP}$ in the vicinity of $m_{Z^0}/2$. 
The points in the left plot in Fig.~\ref{fig:res_Z_limits} at which the 
blue and red dashed lines enter the green region (excluded by LUX
 constraints on the SD cross-section) determine the limiting values 
of $m_{\rm LSP}$ for which the $Z^0$ resonance may give the 
correct relic density of the LSP. 
To be more accurate we used the results for $\Omega h^2$ obtained from 
\texttt{MicrOMEGAs} and found the limiting LSP masses to be approximately
41 and 46.5 GeV.\footnote{These 
results were obtained for the case 
of $\Omega h^2=0.12$ but they do not change much when the uncertainty 
in the calculation of the relic abundance is taken into account.} 
Substituting the corresponding values of 
$|N_{13}^2-N_{14}^2|$ into eq.~\eqref{N13sqmN14sq_0} we find the following 
linear dependence:
\begin{align}
\label{tanb_lam_1}
\tan\beta&\approx 42\times\lambda\quad {\rm for}\;\, 
m_\chi\approx 41\;\,{\rm GeV}\,,\\
\label{tanb_lam_2}
\tan\beta&\approx 44\times\lambda\quad {\rm for}\;\, 
m_\chi\approx 46.5\;\,{\rm GeV}\,.
\end{align}
The values of $\tan\beta$ necessary to obtain good relic abundance 
of the LSP become larger when moving to values of $m_{\rm LSP}$ closer to the 
peak of the resonance (which is slightly below $m_{Z^0}/2$). The situation is 
illustrated in the left plot of Fig.~\ref{fig:channel_Z_om_0}.
One can see that in the region allowed by LUX large $\tan\beta$ is required 
unless $\lambda$ is small ($\sim0.1$ in this case). 
Of course the above limits may become stronger (i.e.~for a given $\lambda$ 
larger $\tan\beta$ might be required) when SD direct detection experiments 
(especially based on the interactions with neutrons) gain better precision. 
For instance, the LZ experiment will be able to explore the entire region 
considered here -- see the left plot in Fig.~\ref{fig:res_Z_limits}. 

In the LSP mass range between the $W^+W^-$ and $t\bar{t}$ thresholds the 
annihilation cross section is dominated by gauge boson ($W^+W^-/Z^0Z^0$) 
final states with the chargino/neutralino exchanged in the t channel. 
The related couplings are proportional to the higgsino components of the 
LSP (the gauginos are decoupled) which for the blind spot \eqref{bs_h_0} 
are related to the LSP-$Z^0$ coupling by eqs.~\eqref{N13_0} and 
\eqref{N14_0}.
The values of $N_{14}$ ($N_{13}$ is smaller by factor $1/\tan\beta$) 
necessary to get $\Omega h^2\sim0.12$ lead to too large 
$\sigma_{\rm SD}$ and are excluded by both LUX and IceCube data
(see Fig.~\ref{fig:cont_om_0}). Thus, the LSP masses in the range 
$m_{W^+} \lesssim m_{\rm LSP} \lesssim m_{t}$ are excluded. 
The only way to have correct relic
abundance consistent with all experimental constraints is to go to very small values of $\lambda$ in order to suppress SI cross-section below the
LUX constraint also away from the blind spot \eqref{bs_h_0} and increase $\tan\beta$ such that the Higgs mass constraint is 
fulfilled.\footnote{
Note that a well-tempered bino-higgsino LSP in MSSM with the mass between the 
$W$ and $t$ masses cannot accommodate all constraints since in that
case the mixing in the neutralino sector is controlled by the gauge coupling constant which is fixed by experiment, see
e.g.~Ref.~\cite{underabundant,welltemperedBOS}.}
More flexibility in the parameter space may appear
if some additional particles are exchanged and/or appear in the final 
state of the LSP annihilation (such situations will be discussed in 
the next sections).

For $m_{\rm LSP}\gtrsim 160$ GeV annihilation into $t\bar{t}$ (via s-channel $Z^0$ exchange) starts to be 
kinematically accessible so smaller higgsino component suffices to have large enough annihilation cross-section to fit $\Omega h^2\approx0.12$. In
consequence, $\Omega h^2\approx0.12$ is obtained for somewhat larger $\tan\beta$ than between the $W^+W^-$ and $t\bar{t}$ thresholds and smaller SD
cross-section is predicted. As a result, IceCube \cite{IceCubeNEW} constraints are satisfied for $m_{\rm LSP}\gtrsim 175$ GeV. However, the new LUX
constraints exclude $m_{\rm LSP}$ up to about 300 GeV for $\Omega h^2=0.12$. 
This lower bound on $m_{\rm LSP}$ may change by about 50~GeV when the 
uncertainties in the calculation of the relic abundance are taken into account. 
It becomes stronger (weaker) for smaller (bigger) values of $\Omega h^2$.
We should also emphasize that the lower bound on the LSP mass from SD 
constraints are the same for the whole class of singlet-doublet fermion DM 
as long as it annihilates dominantly to $t\bar{t}$. In particular, similar 
lower bound of 300 GeV on the LSP mass was recently set by LUX on the 
well-tempered neutralino in MSSM \cite{welltemperedBOS}. 
One can also see in Fig.~\ref{fig:res_Z_limits} that the correct
relic abundance requires 
$|N_{13}^2-N_{14}^2|\sim 0.05$ which, depending on $\tan\beta$, translates to 
$N_{15}^2\sim 0.9-0.95$ (see eq.~\eqref{N15_0}) and such values may be 
explored by XENON1T. 
The right panel of Fig.~\ref{fig:channel_Z_om_0} shows values of $\tan\beta$ 
necessary to get $\Omega h^2=0.12$ as a function of $m_{\rm LSP}$ 
and $\lambda$. Contrary to the $Z^0$ resonance case small values of $\tan\beta$ 
are preferred and hence moderate or large $\lambda$ (in order to have 
big enough Higgs mass at the tree level). 
However, too small values of $\tan\beta$ lead, for a given big value of 
$\lambda$, to a Landau pole below the GUT scale. Thus, the assumption of 
perturbativity up to the GUT scale and the requirement $\Omega h^2=0.12$
give constraints which result in a $\lambda$-dependent upper bound on the 
mass of the LSP. For example, $m_{\rm LSP}\lesssim700$~GeV 
for $\lambda=0.7$ (see Fig.~\ref{fig:channel_Z_om_0}) 
and $m_{\rm LSP}\lesssim800$~GeV for $\lambda=0.6$. 
Let us also note that for large LSP masses coannihilation becomes non-negligible. This effect relaxes the upper bound on $\tan\beta$ and is
increasingly important as $\lambda$ decreases, as can be seen  in Fig.~\ref{fig:channel_Z_om_0} from comparison of full result by {\tt MicrOMEGAs} and
the approximated one with only $t\bar{t}$ included.

Let us comment on two features of the $\Omega h^2\approx0.12$ curves in 
Fig.~\ref{fig:cont_om_0}. First: there are no signs of a resonant 
annihilation with the $h$ boson exchange in the s-channel. 
This is simply the consequence of the blind spot condition 
leading to vanishing (or at least very small) LSP-Higgs coupling. 
This is characteristic of all blind spots without interference effects. 
Second: $\tan\beta$ decreases with $m_\chi$ for all $m_\chi>m_{Z^0}/2$ 
with the exception of the vicinity of the $t\bar{t}$ threshold. 
This is related to the fact that the annihilation cross section 
is directly ($Z^0$ in the s channel) or indirectly (the $VV$ final states) 
connected to the value of $(N_{13}^2-N_{14}^2)$ given by 
eq.~\eqref{N13sqmN14sq_0}. The r.h.s.\ of \eqref{N13sqmN14sq_0} is 
a decreasing function of $m_\chi$ and decreasing function of $\tan\beta$ 
(for $\tan\beta>1+\lambda v_h/m_\chi$). Thus, in order to keep it approximately 
unchanged the increase of $m_\chi$ must be compensated by the decrease 
of $\tan\beta$ (other parameters determining the annihilation cross section 
may change this simple relation only close to the $t\bar{t}$ threshold 
and below the $Z^0$ resonance).

Another comment refers to constraints obtained from
the indirect detection experiments. 
The IceCube upper bounds on $\sigma_{\rm SD}$ change by orders of magnitude 
depending on what channels dominate the LSP annihilation. This can be already 
seen in the simple case discussed in this subsection. The lower bound on 
$\tan\beta$ obtained from the IC data (as a function of $m_{\rm LSP}$) 
visible in Fig.~\ref{fig:cont_om_0} 
drops substantially above the $t\bar{t}$ threshold because the $t\bar{t}$ 
pairs give softer neutrinos as compared to the $W^+W^-$ pairs 
\cite{IceCubeNEW}. Moreover, the latest LUX results on $\sigma_{\rm SD}$ 
lead to stronger bounds in almost all cases. Only for quite heavy LSP
the IC limits are marginally stronger, as may be seen 
in the right panel of Fig.~\ref{fig:res_Z_limits}.

To sum up, in this section we identified two crucial mechanisms 
($Z^0$ resonance and annihilation into $t\bar{t}$) which may give correct 
relic density and are allowed by the experiments. However, both of them rely 
on the $Z^0$ boson exchange in the $s$ channel and therefore are proportional 
to the LSP-$Z^0$ coupling, which controls also the SD cross section of the 
LSP scattering on nucleons. Therefore, the future bounds on such interaction 
will be crucial in order to constrain the parameter space. In fact, XENON1T  is expected to entirely probe regions of the parameter space in
which annihilation into $t\bar{t}$ dominates while LZ will be able to explore the entire region of $Z^0$ resonance.
It is also worth 
noting that the situation presented in Fig.~\ref{fig:cont_om_0} may change 
if we consider light pseudoscalar $a$ with mass $m_a\sim 2m_\chi$. 
Such resonance for singlino-dominated LSP (we require $\kappa\not=0$) is 
controlled mainly by the mixing in (pseudo)scalar sector and hence may not 
be so strongly limited by the SD direct detection experiments. For instance, 
we checked with \texttt{MicrOMEGAs} that for $m_a$ in a few hundred GeV 
range we can easily obtain points in parameter space with correct relic 
density and $\sigma_{\rm SD}$
below the future precision of the LZ 
experiment. In principle, the effect of light pseudoscalar may be also 
important for $2m_\chi\gtrsim m_h+m_a$ when the LSP starts to annihilate 
into $h a$ state which, depending on $\kappa$, may suppress the $t\bar{t}$ 
channel and may weaken the IceCube limits. 
However, in the case considered in this subsection, the contribution 
from the $h a$ channel may be important only for large mixing in the  
pseudoscalar sector. This requires quite large values of $A_\lambda$ 
which leads to unacceptably small values of the Higgs mass.
We will come to these points in the next sections where the annihilation 
channels involving the singlet-dominated pseudoscalar may play 
a more important role.

\subsection{With scalar mixing \label{ssec:bs_fh_mix}}

Next we consider the case when the mixing among scalars is not negligible 
and affects the blind spot condition~\eqref{bs_h_0}, which is now of the form: 
\begin{equation}
\label{bs_fh_mix_eta}
\frac{\tilde{S}_{h\hat{s}}}{\tilde{S}_{h\hat{h}}}
\equiv\gamma
\approx-\eta\,,
\end{equation}
where $\eta$, defined by
\begin{equation}
\label{eta_def}
\eta
\equiv
\frac{N_{15}(N_{13}\sin\beta+N_{14}\cos\beta)}
{N_{13}N_{14}-\frac{\kappa}{\lambda}N_{15}^2}
\,,
\end{equation}
depends on the LSP composition (some formulae expressing $\eta$ in terms 
of the model parameters may be found in Appendix \ref{App:cross-sections}).
In equation \eqref{bs_fh_mix_eta} we introduced also parameter $\gamma$
describing the mixing of the SM-like Higgs with the singlet scalar. 
This mixing can be expressed (for $m_s\gg m_h$ assumed in this section) 
in terms of the NMSSM parameters as:
\begin{equation}
\label{eq:Ssh}
\frac{\tilde{S}_{h\hat{s}}}{\tilde{S}_{h\hat{h}}}
\approx
\lambda v\frac{\left(A_{\lambda}+\langle\pa_S^2 f\rangle\right)\sin2\beta-2\mu}{m_s^2}
\approx
{\rm sgn}\left(\left(A_{\lambda}+\langle\pa_S^2 f\rangle\right)\sin2\beta-2\mu\right)\frac{\sqrt{2|\Delta_{\rm mix}|m_h}}{m_s}
\,,
\end{equation}
where $\Delta_{\rm mix} \equiv m_h - \hat{M}_{hh}$ is the shift of the 
SM-like Higgs mass due to the mixing \cite{Badziak:2013bda}. For $m_s>m_h$ this shift is 
negative so we prefer it has rather small absolute value.

The scenario of higgsino-dominated LSP with $\Omega h^2\approx0.12$ is very similar 
to the analogous case in the MSSM model and requires $|\mu|\approx1.2$ TeV. 
Even the present results from the direct and indirect detection 
experiments constrain possible singlino admixture in the higgsino-dominated 
LSP to be at most of order 0.1. So small singlino component leads to 
negligible changes of $\mu$ necessary to get the observed relic 
density of DM particles.\footnote{
In fact, bigger changes of $\mu$ come from not totally decoupled gauginos,
e.g.~for $M_1$, $M_2$ of order 5 TeV.
}

\begin{figure}[t]
\center
\includegraphics[width=0.49\textwidth]{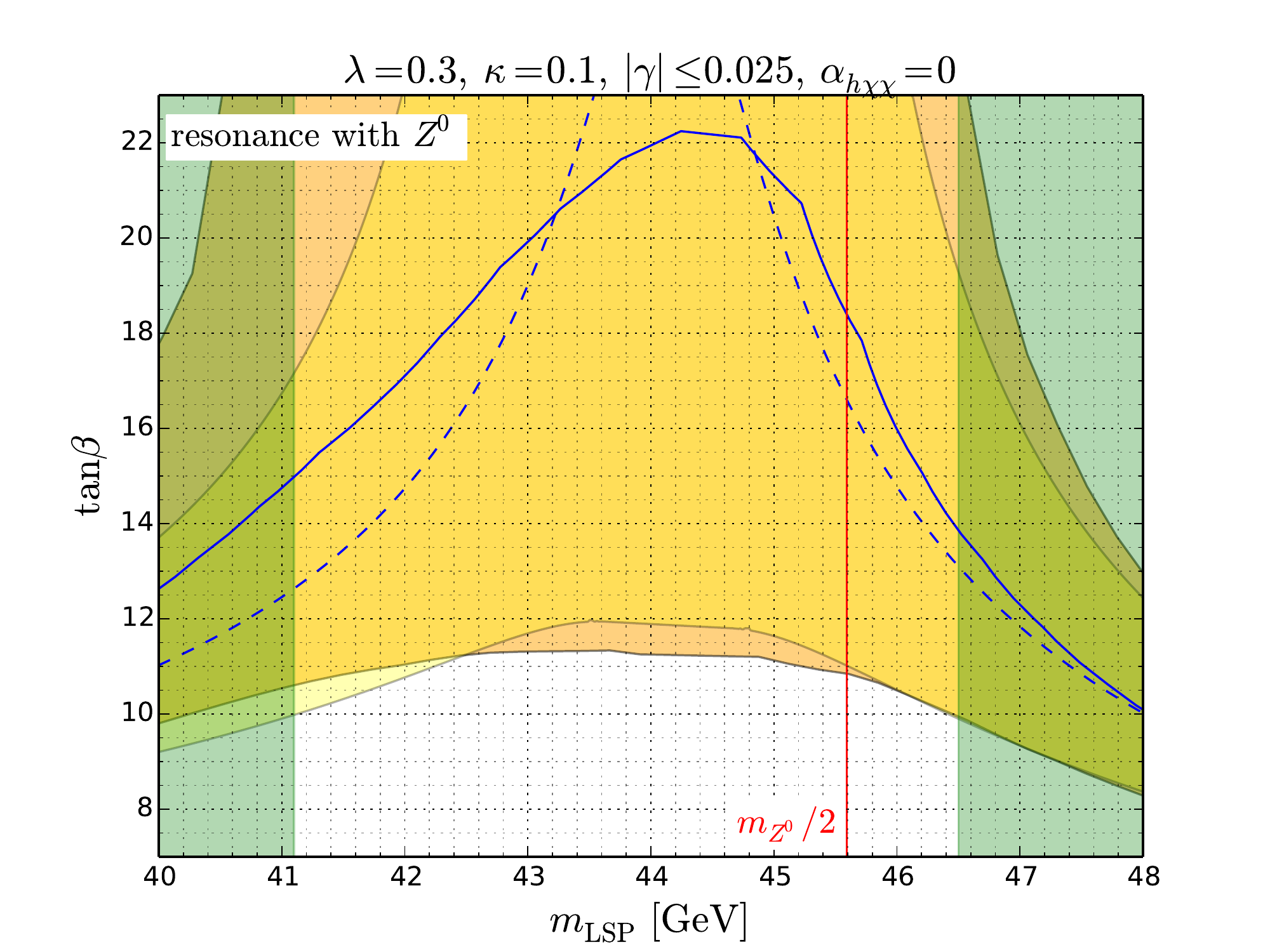}
\includegraphics[width=0.49\textwidth]{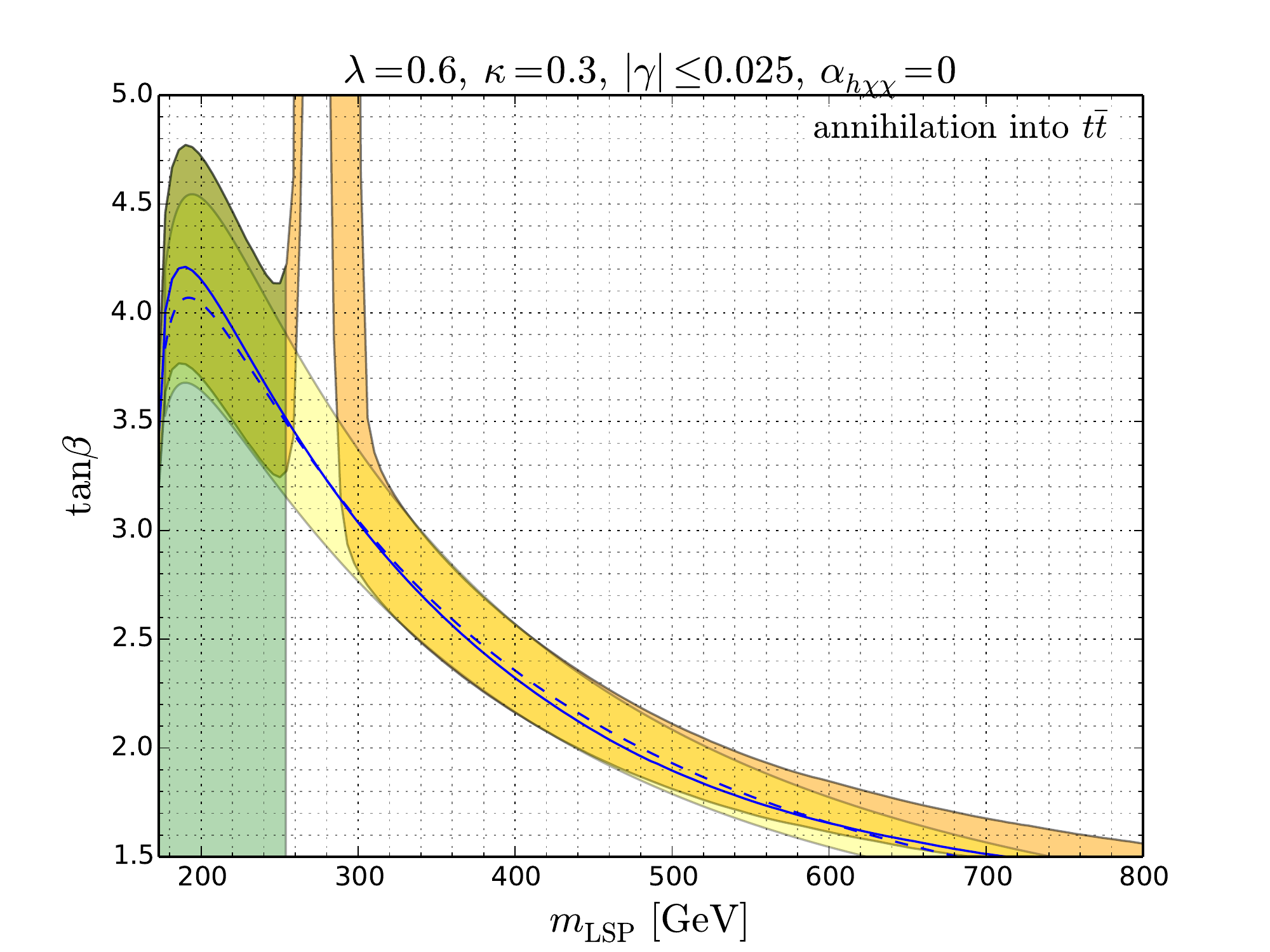}
\caption{The same as in Fig.~\ref{fig:channel_Z_om_0} but now the regions 
with non-zero $s$-$h$ mixing are depicted by colored areas: orange (yellow) 
corresponds to the results from \texttt{MicrOMEGAs} 
(from eqs.~\eqref{eq:res_Z_omega} and \eqref{bs_fh_mix_sing}). We took $m_{\chi}>0$, $\mu>0$, $m_s=600$~GeV and $|\gamma|\leq 0.025$ (corresponding to
 $|\Delta_{\rm mix}|\lesssim 1$ GeV), where $\gamma<0$ ($>0$) refers to the lower (upper) limit of a given area.  
The shift of the orange region with respect to the yellow one for LSP 
masses above about $m_s$ is due to the annihilation channels containing 
$s$ in the final state and co-annihilation effects which become more important for larger LSP masses.
}
\label{fig:channel_Z_om_mix}
\end{figure}

Thus, similarly as before, we focus on SI blind spots with 
$\Omega h^2\approx0.12$ 
for singlino-dominated LSP. In this case and for non-negligible $\kappa$ 
(more precisely, when $|N_{13}N_{14}|\ll\frac{|\kappa|}{\lambda}N_{15}^2$) 
the blind spot condition may be approximated by:
\begin{equation}
\label{bs_fh_mix_sing}
\frac{m_\chi}{\mu}-\sin2\beta\approx\gamma\,\frac{\kappa}{\lambda}\,
\frac{\mu}{\lambda v_h}
\left(1-\left(\frac{m_\chi}{\mu}\right)^2\right)\,.
\end{equation}
This condition is a quadratic equation for $\mu$ and has solutions 
only when
\begin{equation}
\cos^2\beta>\frac12
-\left|\frac12+\gamma\,\frac{\kappa}{\lambda}\,\frac{m_\chi}{\lambda v_h}\right|
\,.
\label{gamma_BS_fh_mix}
\end{equation}
One can see from \eqref{bs_fh_mix_sing} that for $\gamma\kappa\mu>0$ we have 
always $m_\chi\mu>0$ and the LSP has more higgsino fraction than when 
condition~\eqref{bs_h_0} holds.
In the opposite case i.e.~$\gamma\kappa\mu<0$ 
we can have either $m_\chi\mu>0$ with slightly smaller higgsino fraction or 
strongly singlino-dominated LSP with $m_\chi\mu<0$. However, for values of $|\gamma|$ small enough not to induce large negative $\Delta_{\rm mix}$
the higgsino component of the LSP with $m_\chi\mu<0$ is too small to obtain $\Omega h^2\approx 0.12$. Therefore, from now on we focus on the case
$m_\chi\mu>0$. 
Solving eq.~\eqref{bs_fh_mix_sing} for the ratio $m_\chi/\mu$ and substituting 
the solution into eq.~\eqref{N13sqmN14sq_0} one can find the difference 
of two higgsino components of the LSP. For small $|\gamma|$ it can be 
approximated as \eqref{N13sqmN14sq_0} with a small correction:
\begin{align}
\label{N13sqmN14sq_mix}
\left|N_{13}^2-N_{14}^2\right|
\approx
&\left(\frac{\lambda v_h}{m_\chi}\right)^2
\frac{\sin^22\beta}{|\cos2\beta|}
\left[1+\left(\frac{\lambda v_h}{m_\chi}\right)^2\tan^22\beta\right]^{-1}
\nonumber\\
&+
\gamma\,\frac{\kappa}{\lambda}\,\frac{\lambda v_h}{m_\chi}\,
\frac{2}{|\cos2\beta|^3}
\left[1-\left(\frac{\lambda v_h}{m_\chi}\right)^2
\frac{\sin^42\beta}{\cos^22\beta}\right]
\left[1+\left(\frac{\lambda v_h}{m_\chi}\right)^2\tan^22\beta\right]^{-2}
.
\end{align}
As we already mentioned discussing eq.~\eqref{N13sqmN14sq_0}, the first term 
in the r.h.s.\ of the last equation is a decreasing function of $\tan\beta$
(unless $\tan\beta$ is very close to 1). So, in order to keep the value 
of $\left|N_{13}^2-N_{14}^2\right|$ necessary to get $\Omega h^2\approx0.12$, 
the contribution from the second line of \eqref{N13sqmN14sq_mix} may be 
compensated by increasing (decreasing) the value of $\tan\beta$ 
when $\gamma\kappa\mu>0$ ($<0$).\footnote{\label{foot_signs}
The sign of the second term in the r.h.s.~of~\eqref{N13sqmN14sq_mix} 
is determined by the sign of product $\gamma\kappa m_\chi$ 
but for considered here case of $m_\chi\mu>0$ it is the same as that 
of $\gamma\kappa\mu$.}
This effect from the $s$-$h$ mixing is illustrated in 
Fig.~\ref{fig:channel_Z_om_mix}. One can see that indeed, depending on the 
sign of $\gamma\kappa\mu$, we can have smaller or larger (in comparison to 
eq.~\eqref{bs_h_0}) values of $\tan\beta$ for a given LSP mass, while keeping 
$\Omega h^2\approx 0.12$. In particular, non-negligible Higgs-singlet mixing may relax the upper bound on the LSP mass arising from the
perturbativity up to the GUT scale.
It is also important to emphasize that the resonant annihilation
with $s$ scalar exchanged in the s channel is quite generic for this kind 
of blind spot (see the right plot in Fig.~\ref{fig:channel_Z_om_mix}, 
where we have chosen $m_s=600$~GeV) because it is easier to have substantial 
 $s$-$h$ mixing when the singlet-dominated scalar is not very heavy. Moreover, the presence of resonant annihilation via $s$ exchange can relax the
lower limit on the LSP mass from LUX constraints on the SD cross-section, as seen in the right panel of Fig.~\ref{fig:channel_Z_om_mix}, because in
such a case smaller higgsino component is required to obtain correct relic density.

\section{Blind spots  and relic density with light singlets
\label{sec:bs_fhs}}

\vspace{2ex}
\noindent
Now we move to the case when the singlet-dominated scalar is lighter than 
the SM Higgs. Neglecting the effect from the heavy doublet $H$ exchange 
for the SI cross section (i.e.\ setting $f_H$ to zero) 
the blind spot condition may be written in the following form~\cite{BS_NMSSM}:
\begin{equation}
\label{bs_fhs_mix}
\frac{\gamma+\mathcal{A}_s}{1-\gamma\mathcal{A}_s}=-\eta\,,
\end{equation}
where
\begin{equation}
\label{eq:As}
\mathcal{A}_s\approx
-\gamma\,\frac{1+c_s}{1+c_h}\left(\frac{m_h}{m_s}\right)^2
\end{equation}
and parameters
\begin{equation}
\label{c}
c_{h_i}\equiv1+\frac{\tilde{S}_{{h_i}\hat{H}}}{\tilde{S}_{{h_i}\hat{h}}}
\left(\tan\beta-\frac{1}{\tan\beta}\right)
\end{equation}
measure (in the large $\tan\beta$ limit) the ratio of the couplings, 
normalized to the SM values, of the $h_i$ $(=s,h)$ scalar to the $b$ quarks 
and to the $Z^0$ bosons.

\begin{figure}[t]
\center
\includegraphics[width=0.49\textwidth]{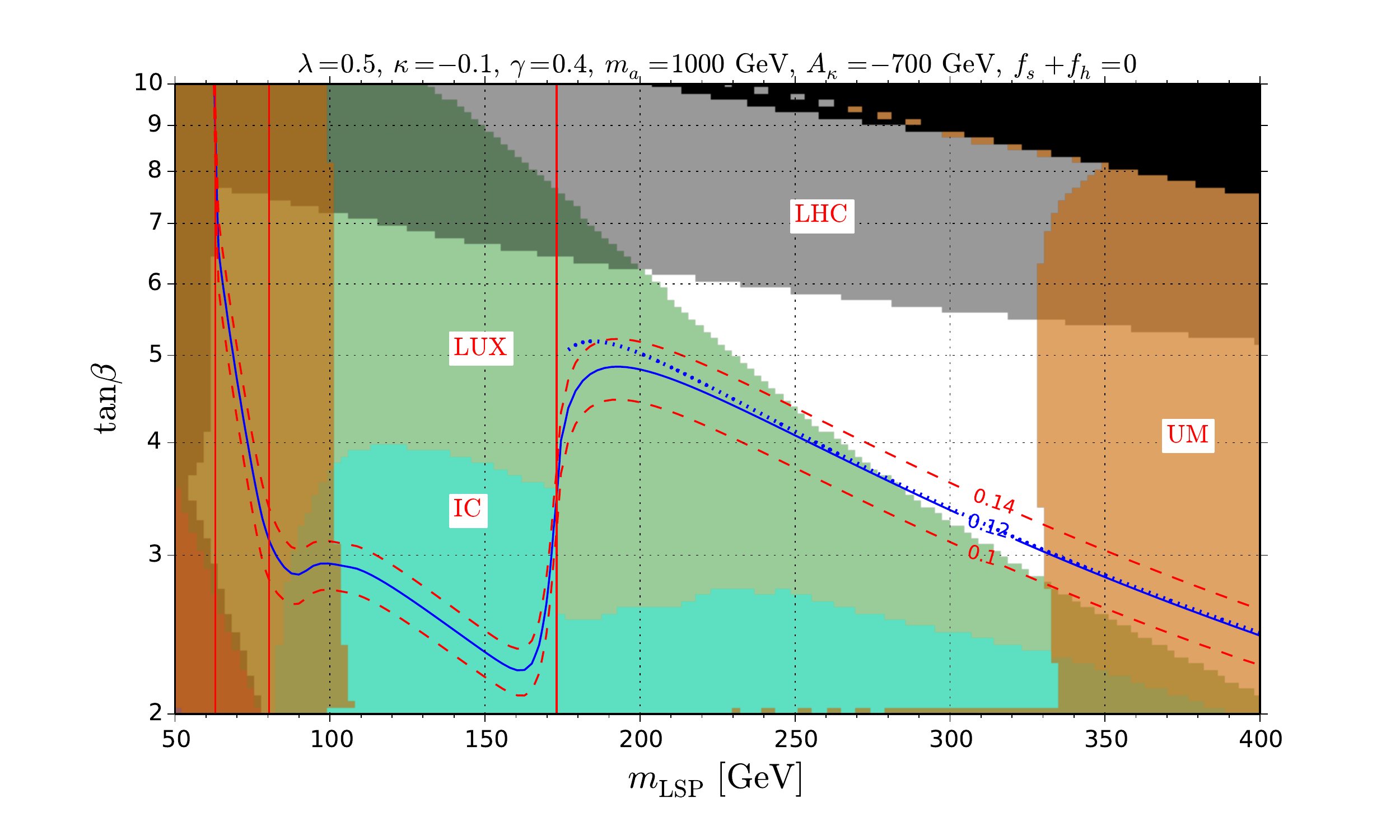}\hspace{-4ex}
\includegraphics[width=0.49\textwidth]{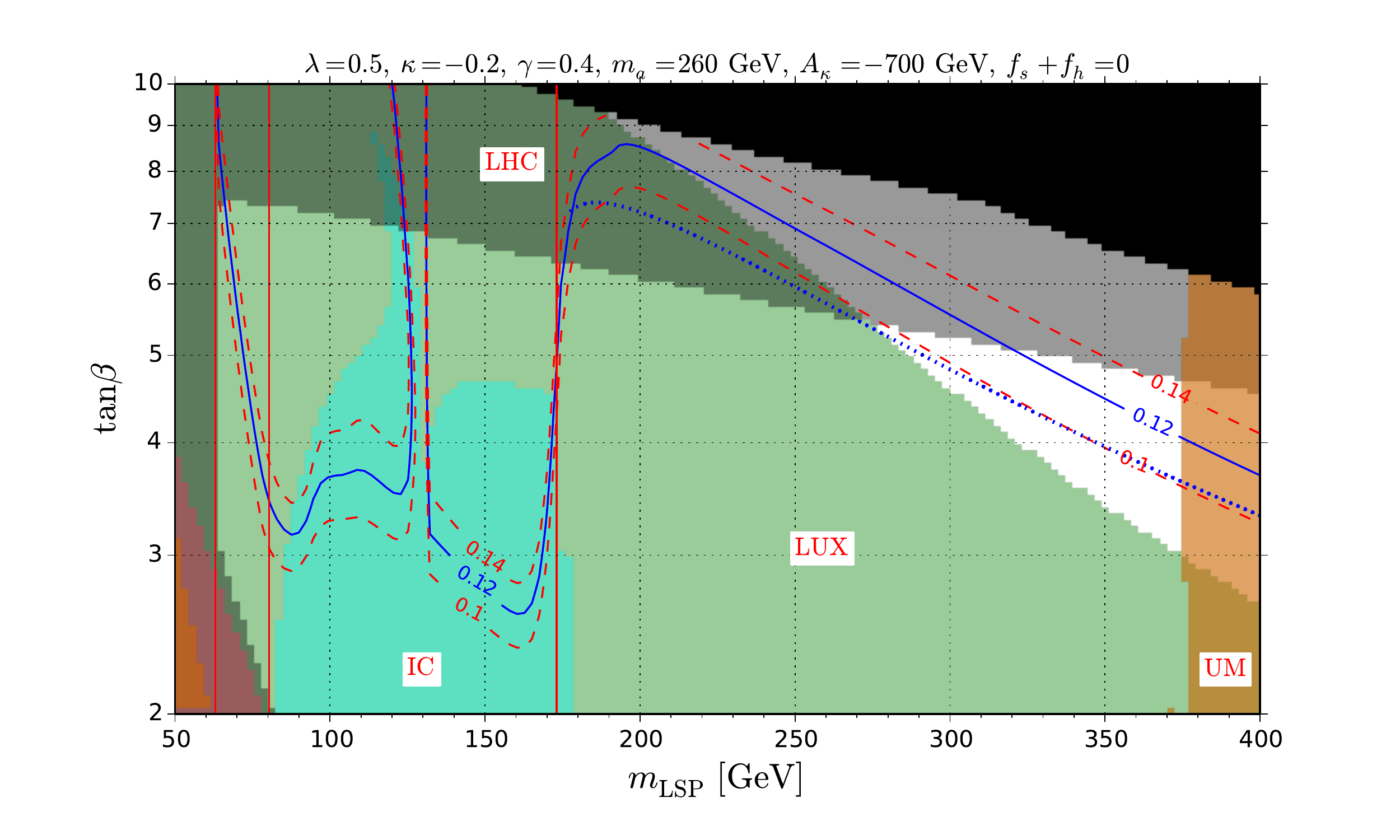}
\includegraphics[width=0.49\textwidth]{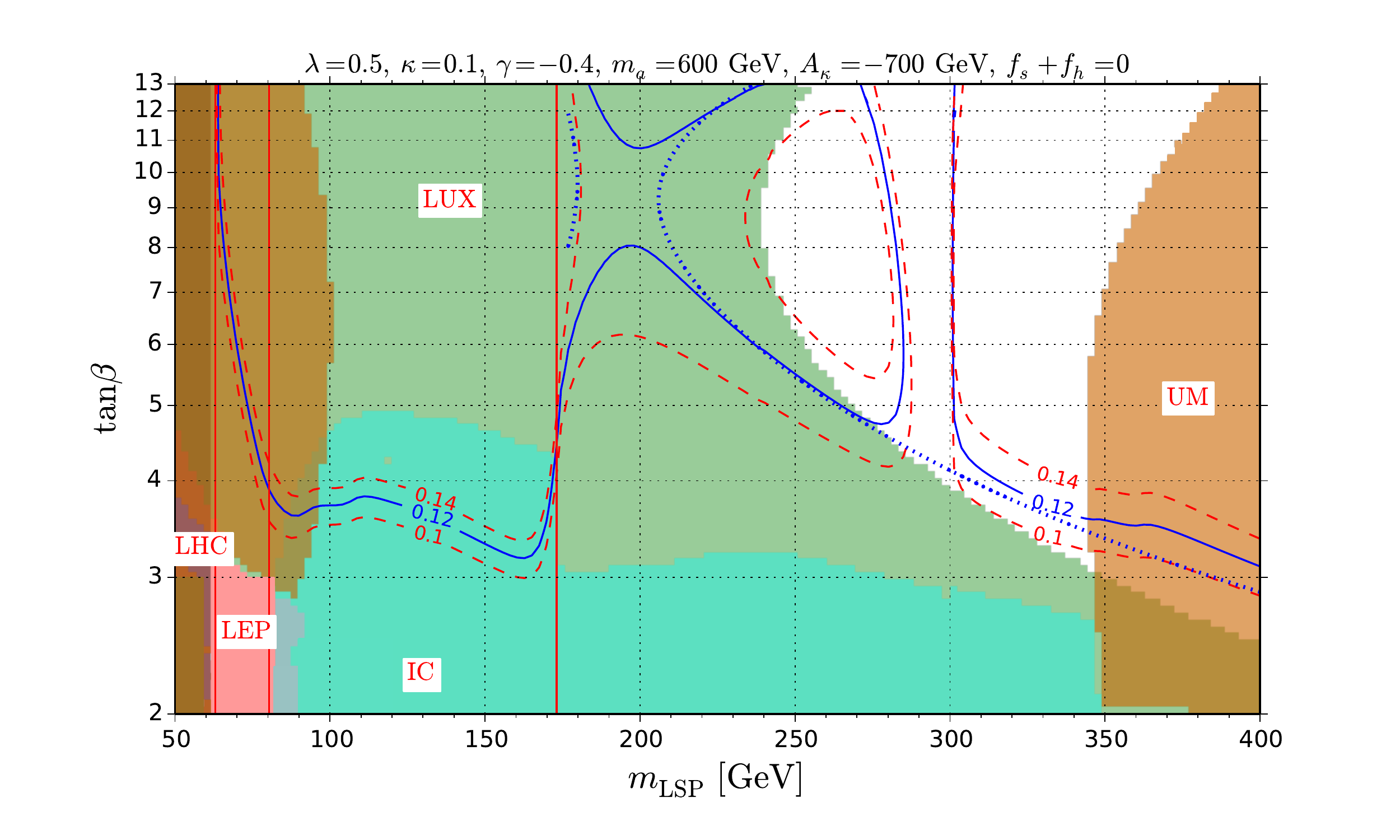}\hspace{-4ex}
\includegraphics[width=0.49\textwidth]{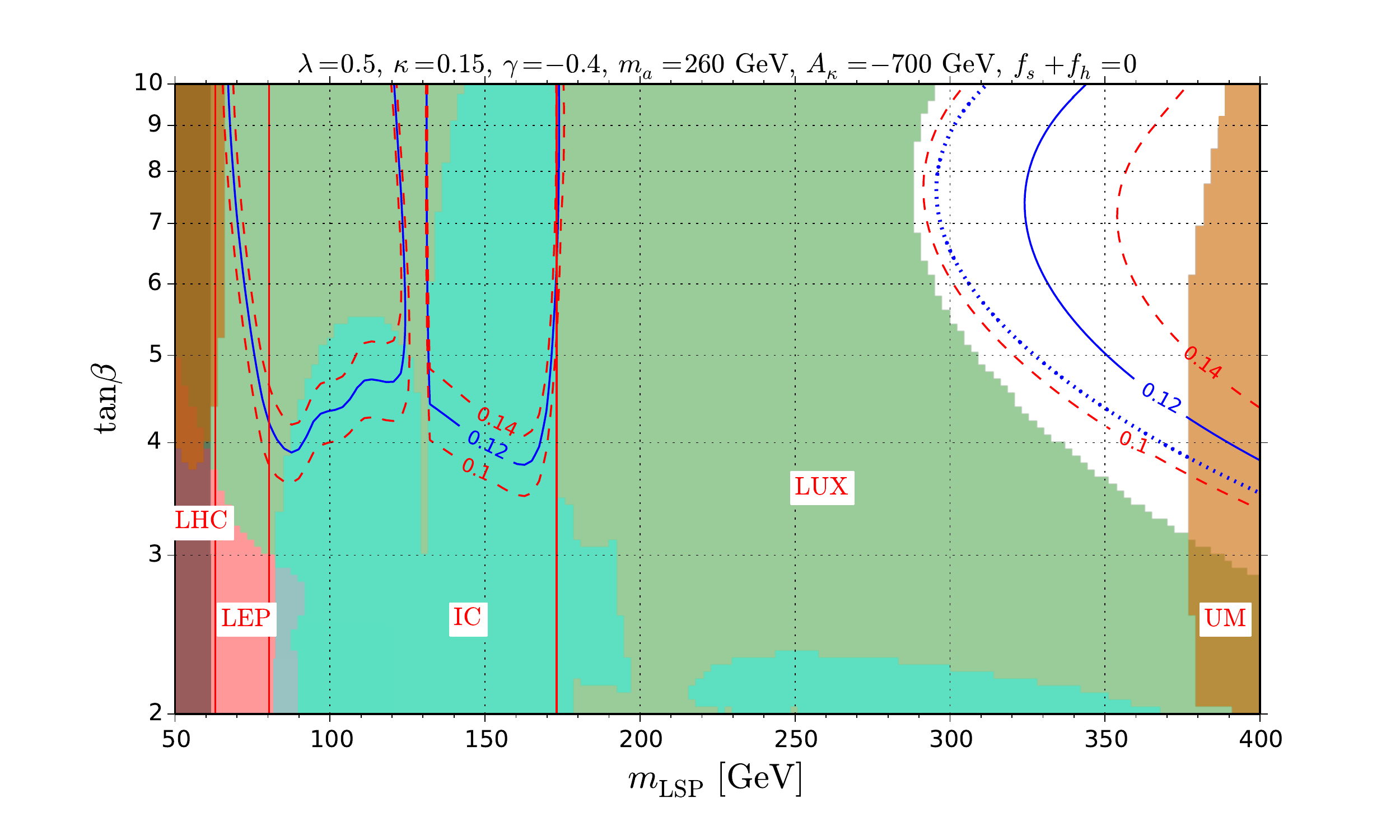}
\caption{The same as in Fig.~\ref{fig:cont_om_0} but for $m_s<m_h$. 
We took $m_{\chi}>0$, $\mu>0$, $m_H=4$~TeV, $m_s=95$~GeV  and $|\gamma|=0.4$, which corresponds to $\Delta_{\rm mix}\sim 4$~GeV. 
In brown areas {\tt NMSSMTools} reports unphysical global minimum (UM) while in grey ones the LHC constraints on the Higgs production and decay are
violated (LHC). Black
regions in the upper right corners denote the points where the condition~\eqref{gamma_BS_fhs} does not hold.
}
\label{fig:cont_om_fhs}
\end{figure}

In the rest of this section we will consider singlino-like LSP, because 
the case of higgsino-dominated LSP does not differ much from the one 
described in section~\ref{ssec:bs_fh_mix}. The blind spot condition \eqref{bs_fhs_mix},
analogously to a simpler case \eqref{bs_fh_mix_eta}, may be 
approximated by a quadratic equation for $\mu$ 
\begin{equation}
\label{bs_fhs}
\frac{m_\chi}{\mu}-\sin2\beta
\approx
\frac{\gamma+\mathcal{A}_s}{1-\gamma\mathcal{A}_s}\,
\frac{\kappa}{\lambda}\,
\frac{\mu}{\lambda v_h}
\left(1-\left(\frac{m_\chi}{\mu}\right)^2\right)\
\end{equation}
which has solutions only if
\begin{equation}
\label{gamma_BS_fhs}
\cos^2\beta
>
\frac12-\left|\frac12
+
\gamma\,\frac{\kappa}{\lambda} \,\frac{m_\chi}{\lambda v_h}
\,\frac{1-\frac{1+c_s}{1+c_h}\left(\frac{m_h}{m_s}\right)^2}
{1+\gamma^2\,\frac{1+c_s}{1+c_h}\left(\frac{m_h}{m_s}\right)^2}
\right|.
\end{equation}
The last condition may be interpreted as the upper bound on $\tan\beta$ 
(lower bound on $\cos\beta$). It is nontrivial when its r.h.s.\ is 
positive i.e.\ when the second term under the absolute value is negative 
but bigger than $-1$. The bound is strongest when that term 
equals $-\frac12$. However, usually the absolute value of that term is 
smaller then $\frac12$ because the $h$-$s$ mixing measured by 
$\gamma$ is rather small. So, typically the bound on $\tan\beta$ 
becomes stronger with increasing LSP mass or increasing $|\kappa|$ 
(with other parameters fixed). 
We focus again on the more interesting case $m_\chi\mu>0$ because for $m_\chi\mu<0$ the blind spot condition may be satisfied only for the LSP
strongly dominated by singlino which typically leads to too large relic density.\footnote{For $m_\chi\mu<0$, $\Omega h^2\approx0.12$ may be obtained
only when resonant annihilation via singlet (pseudo)scalar exchange is dominant (we describe this phenomenon for $m_\chi\mu>0$) or $\lambda\ll\kappa$
which is less interesting from the point of view of the Higgs mass. }
Since in this section we consider $m_s<m_h$, $|\mathcal{A}_s|$ is typically  larger than $|\gamma|$
(unless $c_s$ and/or $c_h$ deviate much from 1, which under some conditions may happen which we discuss in more detail
later in this section)~-- in such a case the condition~\eqref{gamma_BS_fhs} is always fulfilled for $\gamma\kappa\mu<0$ (see
Fig.~\ref{fig:cont_om_fhs}), whereas for $\gamma\kappa\mu>0$ 
(see comment in footnote~\ref{foot_signs} ) 
there is an upper bound on $\tan\beta$ which gets stronger for larger
values of $\gamma\kappa$.

In the following discussion we focus on big values of $|\gamma|$ because 
they lead to a relatively big positive contribution to the Higgs mass from 
the Higgs-singlet mixing \cite{Badziak:2013bda}. In our numerical analysis 
we take $|\gamma|=0.4$ which corresponds to $\Delta_{\rm mix}\sim 4$~GeV. 
In order to emphasize new features related to the modification of the BS 
condition we also consider rather large values of 
$|\kappa|\sim\mathcal{O}(0.1)$. For such choices of the parameters there 
are no viable blind spots for $\gamma\kappa\mu>0$ in accord with the  
discussion above so we focus on $\gamma\kappa\mu<0$.

The fact that our blind spot condition now comes from destructive interference 
between $f_h$ and $f_s$ amplitudes (rather than vanishing Higgs-LSP coupling) 
influences strongly the relation between the DM relic density, especially 
in small LSP mass regime, and other experimental constraints. 
Now the Higgs-LSP coupling is not negligible so the LSP mass below $m_h/2$ 
is forbidden, or at least very strongly disfavored, by the existing bounds 
on the invisible Higgs decays \cite{Higgs_inv_exp}. Thus, resonant annihilation with $Z^0$ or $s$ 
(in this section we consider $s$ lighter than $h$) exchanged in the s channel 
can not be used to obtain small enough singlino-like LSP relic abundance. 
As concerning the $h$ resonance: it may be used but only the ``half'' of it 
with $m_{\rm LSP}\gtrsim m_h/2$ (this effect is visible in all panels of 
Fig.~\ref{fig:cont_om_fhs}). However, we found that other experimental constraints, such as the ones from the LHC and/or LUX
exclude even this ``half'' of the $h$ resonance when the mixing parameter 
$|\gamma|$ is large.

In general NMSSM the masses of singlet-like scalar $s$ and pseudoscalar $a$ 
are independent from each other so let us first consider the situation
when $a$ is heavy. The case with $m_a=1$~TeV is presented in the upper left 
panel of Fig.~\ref{fig:cont_om_fhs}. The contours of 
$\Omega h^2=0.10,\,0.12,\,0.14$ above the $t\bar{t}$ threshold are quite 
similar to the case with heavy singlet. The only difference is that now 
somewhat larger values of $\tan\beta$ are preferred but
even in this case they cannot exceed about 5.

Let us now check what happens when the lighter pseudoscalar is also 
singlet-dominated (i.e.~$a_1=a$) and relatively light. The existence 
of such light pseudoscalar is very important for both the relic abundance 
of the LSP and the constraints from the IceCube experiment. 
Let us now discuss these effects in turn.

The DM relic density is influenced by a pseudoscalar in two ways. 
First is given by possible resonant annihilation with $a$ exchanged in the 
s channel. This possibility is interesting only for $m_a\gtrsim m_h$ 
because for lighter $a$ one still has problems with non-standard Higgs decays
constrained by the LHC data 
(see above)\footnote{
The situation may be different if the blind spot condition is of the standard 
form~\eqref{bs_h_0} -- we will come to this point at the end of this 
section.}.
However, as for any narrow resonance, the DM relic density may be in agreement 
with observations only for a quite small range of the DM mass (for 
a given $a$ mass). One can see this in three panels (except the upper left one) 
of Fig.~\ref{fig:cont_om_fhs}. 
Second effect is related to new annihilation final states including 
the singlet-dominated pseudoscalar, namely $sa$, $ha$, $aa$
(in addition to similar channels involving only scalars: $ss$, $sh$, $hh$). 
It is best illustrated in the upper right panel of Fig.~\ref{fig:cont_om_fhs} 
for $m_a=260$~GeV. In this case the $sa$ threshold roughly coincides with 
the $t\bar{t}$ threshold. Near this threshold the curves of constant 
$\Omega h^2\approx0.12$ go up towards bigger values of $\tan\beta$ and 
leave the region excluded by the LUX data for smaller LSP mass 
than in the case with heavy singlets. The reason is quite simple. 
With increasing contribution from the annihilation channels mediated by 
(non-resonant) pseudoscalar exchange smaller contribution from the 
channels mediated by $Z^0$ exchange is enough to get the desired value 
$\Omega h^2\approx0.12$. Moreover, smaller LSP-$Z^0$ coupling is obtained 
for bigger values of $\tan\beta$ so larger values of $\tan\beta$ are 
preferred than in the case with heavy $a$.
As a result the lower possible LSP mass consistent with the LUX SD limits 
is almost 100~GeV smaller when $a$ is relatively light. The precise values 
depend on the relic density and for $\Omega h^2=0.12$ is around 200 GeV 
instead of around 300 GeV as in the case with heavy singlets.

The behavior of the $\Omega h^2=0.12\pm0.02$ curves close to and 
slightly above the $t\bar{t}$ threshold depends on the parameters. 
Particularly important is the sign of $\gamma$. We see from the right
panels of Fig.~\ref{fig:cont_om_fhs} that even for the same mass of the 
pseudoscalar, $m_a=260$ GeV in this case, the plots are very different for
different signs of $\gamma$. Most differences originate from the fact 
that $\gamma>0$ implies $c_h>1$ and $c_s<1$ while for $\gamma<0$ the
inequalities are reversed. There are two important implications of these 
correlations which we describe in the following.

Firstly, the LHC constraints from the Higgs coupling
measurements are stronger for $\gamma>0$ because in such a case the Higgs coupling (normalized to the SM) to bottom quarks
is larger than the one to gauge
bosons. In consequence, the Higgs branching ratios to gauge bosons is suppressed as compared to the SM. Moreover, non-zero $\gamma$ results in
suppressed Higgs production cross-section so if $|\gamma|$ is large enough the Higgs signal strengths in gauge boson decay channels is too small to
accommodate the LHC Higgs data  which agree quite well with the SM prediction. Moreover, a global fit to the current Higgs data shows some
suppression of the Higgs coupling to bottom quarks \cite{Higgscomb}
\footnote{The current fit also indicates an enhancement of the top Yukawa coupling. In NMSSM suppression  of
the bottom Yukawa coupling is correlated with enhancement of the top Yukawa coupling which has been recently studied in Refs.~\cite{tth1,tth2}.}
which disfavors $c_h>1$, hence also large $\gamma>0$. It can be seen from the upper right panel of Fig.~\ref{fig:cont_om_fhs} that for $\gamma=0.4$
the LHC excludes some of the interesting part of the parameter space which is allowed by LUX due to the LSP annihilations into $sa$ final state. The
LHC constraints can be satisfied for small values of $\gamma$ but this comes at a price of smaller $\Delta_{\rm mix}$, hence somewhat heavier stops.

Secondly,  $|\mathcal{A}_s|$ is larger for $\gamma<0$ than in the opposite case (see~eq.~\eqref{eq:As}). Moreover, since deviations of $c_s$ and $c_h$
from 1 grow with $\tan\beta$, $|\mathcal{A}_s|$ increases (decreases) with $\tan\beta$ for negative (positive) $\gamma$. For $\gamma>0$, this implies
that for large enough $\tan\beta$ the r.h.s. of the blind spot condition~\eqref{bs_fhs} changes sign and $\gamma\kappa\mu<0$ is no longer preferred.
Equivalently, the upper bound on $\tan\beta$ in eq.~\eqref{gamma_BS_fhs} gets stronger as $\tan\beta$ grows so it is clear that at some point
condition~\eqref{gamma_BS_fhs} is violated. The appearance of the violation of the blind spot condition is clearly visible at large $\tan\beta$ in
the upper panels of Fig.~\ref{fig:cont_om_fhs} (the black regions). For $\gamma<0$ instead the blind spot condition may be always fulfilled for
$\gamma\kappa\mu<0$ (by taking e.g.~appropriate value 
of $\mu$).\footnote{The 
only exception is when $c_h<-1$ but this may happen only for very large
$\tan\beta$ and/or very light $H$.}
However, there is interesting phenomenon that may happen for $\gamma>0$ above the $t\bar{t}$ threshold which is well visible in the lower right panel
of Fig.~\ref{fig:cont_om_fhs}. The values of $\tan\beta$ corresponding to $\Omega h^2=0.12$ grow rapidly just above the $t\bar{t}$ threshold and the
there is a gap in the LSP masses 
for which there are no solutions with SI BS and observed value of $\Omega$. 
Such solutions appear again for substantially bigger $m_{\rm LSP}$ (above 300 GeV in this case). The reason why $\Omega h^2=0.12$ curve is almost
vertical
near $m_{\rm LSP}\sim m_t$ and the gap appears is related to the fact that $m_\chi/\mu$ varies very slowly with $\tan\beta$, which results in fairly
constant $|N_{13}^2-N_{14}^2|$ which determines the $t\bar{t}$ annihilation cross-section, hence also $\Omega h^2$. The weak dependence of
$m_\chi/\mu$ on $\tan\beta$ originates from the fact that for increasing $\tan\beta$ both sides of the blind spot condition~\eqref{bs_fhs} grow. The
l.h.s. grows because of decreasing $\sin(2\beta)$ while the r.h.s. due to increasing $c_s$. Of course, the fact that these two effects approximately
compensate each other relies on specific choice of parameters and does not necessarily hold e.g.~for different values of $\kappa$.

The presence of light $a$ influences also the IceCube constraints 
in a way depending on the LSP mass. For $m_\chi\gtrsim(m_a+m_s)/2$ 
(assuming $m_a>m_s$) the IceCube constraints are very much relaxed 
and become practically unimportant for the cases discussed in this section. 
This is so because the additional annihilation channels 
(into $sa$, $ha$, $ss$, $sh$) at $v=0$ lead to softer 
neutrinos as compared to otherwise dominant $VV$ channels
(or $t\bar{t}$ channel for even heavier pseudoscalar). 
The situation is different (and more complicated) for LSP masses 
between the $W^+W^-$ and $as$ thresholds. In this region  
one can have destructive/constructive interference between $Z^0$ and $a$-mediated 
amplitudes\footnote{
Note that these are the only non-negligible amplitudes which have non-zero 
$a$ term in $\sigma v$ expansion~-- see~\eqref{sigmaV_expansion}.
} 
for $b\bar{b}$ annihilation at $v=0$ which strengthens/reduces the constraints 
(IceCube limit on SD cross section are two orders of magnitude stronger 
for $VV$ than for $b\bar{b}$). In our case the effect depends on the sign of $\kappa$: 
for $\kappa<0$ the IceCube limits are strengthen (relaxed) for $m_\chi\lesssim m_a$ ($\gtrsim m_a$) and vice versa for $\kappa>0$
-- see the lower panels of Fig.~\ref{fig:cont_om_fhs}.

In the examples considered in this section and shown in 
Fig.~\ref{fig:cont_om_fhs} we have chosen the sign of $\mu$ to be positive 
(then $m_\chi$ is also positive because, as discussed earlier, 
there are no interesting solutions with $m_\chi\mu<0$). The corresponding 
solutions with negative $\mu$, and also changed signs of other parameters 
like $\kappa$, $\gamma$ and $A_\kappa$, are qualitatively quite similar. 
Of course there are some quantitative changes. Contours in the corresponding 
plots are slightly shifted towards smaller or bigger 
(depending on the signs of other parameters) values of $\tan\beta$. 
Typically also the regions of unphysical minima are moved towards 
bigger values of the LSP mass.

As explained in detail in Ref.~\cite{BS_NMSSM}, 
one can also get vanishing spin-independent cross section when the standard 
BS condition~\eqref{bs_h_0} is fulfilled. Then, the scalar sector has no effect 
on the blind spot condition. In such a case, we can have as large 
$\Delta_{\rm mix}$ as is allowed solely by the LEP and LHC constraints, 
irrespective of the DM sector. 
Interestingly, the standard BS condition appears 
also, in a non-trivial way, when $\frac{|\kappa|}{\lambda}$ is relatively 
small (we still consider singlino-dominated LSP) and both terms in the 
denominator of~\eqref{eta_def} are comparable and approximately cancel each 
other. The blind spot condition~\eqref{bs_fh_mix_eta} requires 
$\eta$ to be fixed and not very large. From eq.~\eqref{eta_lambda/mu} 
we see that in such a case a small denominator, for a singlino-dominated LSP, 
may be compensated by small factor $(m_\chi/\mu-\sin2\beta)$ which means 
that the simplest BS condition \eqref{bs_h_0} is approximately fulfilled.
In both cases, the analysis 
performed in subsection~\ref{ssec:bs_fh_nomix} holds. However, it should 
be noted that small $|\kappa|$ weakens the singlet self-interaction 
$\sim\kappa S^3$ and hence may decrease the above-mentioned effects 
from the $a$ exchange.

\section{$\mathbb{Z}_3$-symmetric NMSSM}
\label{sec:Z3}

All the analysis presented till this point apply to a general NMSSM. 
In this section we focus on the most widely studied version of NMSSM with 
$\mathbb{Z}_3$ symmetry. In this model there is no dimensionful 
parameters in the superpotential:
\begin{equation}
\label{W_Z3NMSSM}
 W_{\rm NMSSM}= \lambda SH_uH_d + \kappa S^3/3 \,,
\end{equation}
while the soft SUSY breaking Lagrangian is given by \eqref{Lsoft} with $m_3^2=m_S'^2=\xi_S=0$. This model has five free parameters less than general
NMSSM which implies that some physical parameters important for dark matter sector are correlated. The main features of $\mathbb{Z}_3$-symmetric
NMSSM relevant for phenomenology of neutralino dark matter are summarized below:
\begin{itemize}
\item ${\rm sgn ( m_\chi \mu)}={\rm sgn}(\kappa)$
 \item LSP dominated by singlino implies
 \begin{equation}
 \label{Z3cond1}
  |\kappa| < \frac12\lambda \,.
 \end{equation}
\item Neither singlet-like scalar nor singlet-like pseudoscalar can be 
decoupled due to the following tree-level relation 
(for singlino LSP after taking into account the leading 
contributions from the mixing with both scalars coming from 
the weak doublets, $\hat{h}$ and $\hat{H}$):
\begin{equation}
 \label{Z3cond2}
 m_s^2 + \frac13 m_a^2  \approx  m_{\rm LSP}^2 + \gamma^2(m_s^2 - m_h^2) \,.
\end{equation}
Masses of both singlet-dominated scalar and pseudoscalar are at most 
of order $m_{\rm LSP}$.
\item Phenomenologically viable (small) Higgs-singlet mixing leads to the following  tree-level relation:
\begin{equation}
 \label{Z3cond3}
 M_A\equiv |M_{\hat{A}\hat{A}}| \approx \frac{2 |\mu|}{\sin (2\beta)}\sqrt{1-\frac{\kappa\sin (2\beta)}{2\lambda}} \approx \frac{2
|\mu|}{\sin(2\beta)} \,,
\end{equation}
\end{itemize}
where the last approximation is applicable for large $\tan\beta$ and/or singlino-like LSP and forbids resonant LSP annihilation via heavy Higgs
exchange. Such resonance may be present only for $\lambda\ll1$ since only in such a case significant deviation from relation \eqref{Z3cond3}
is possible. Important constraints on dark matter sector of $\mathbb{Z}_3$-symmetric NMSSM follow from relation \eqref{Z3cond2} which we discuss in
more detail in the following subsections.

\subsection{Heavy singlet scalar}

%
\begin{figure}
\center
\includegraphics[width=0.49\textwidth]{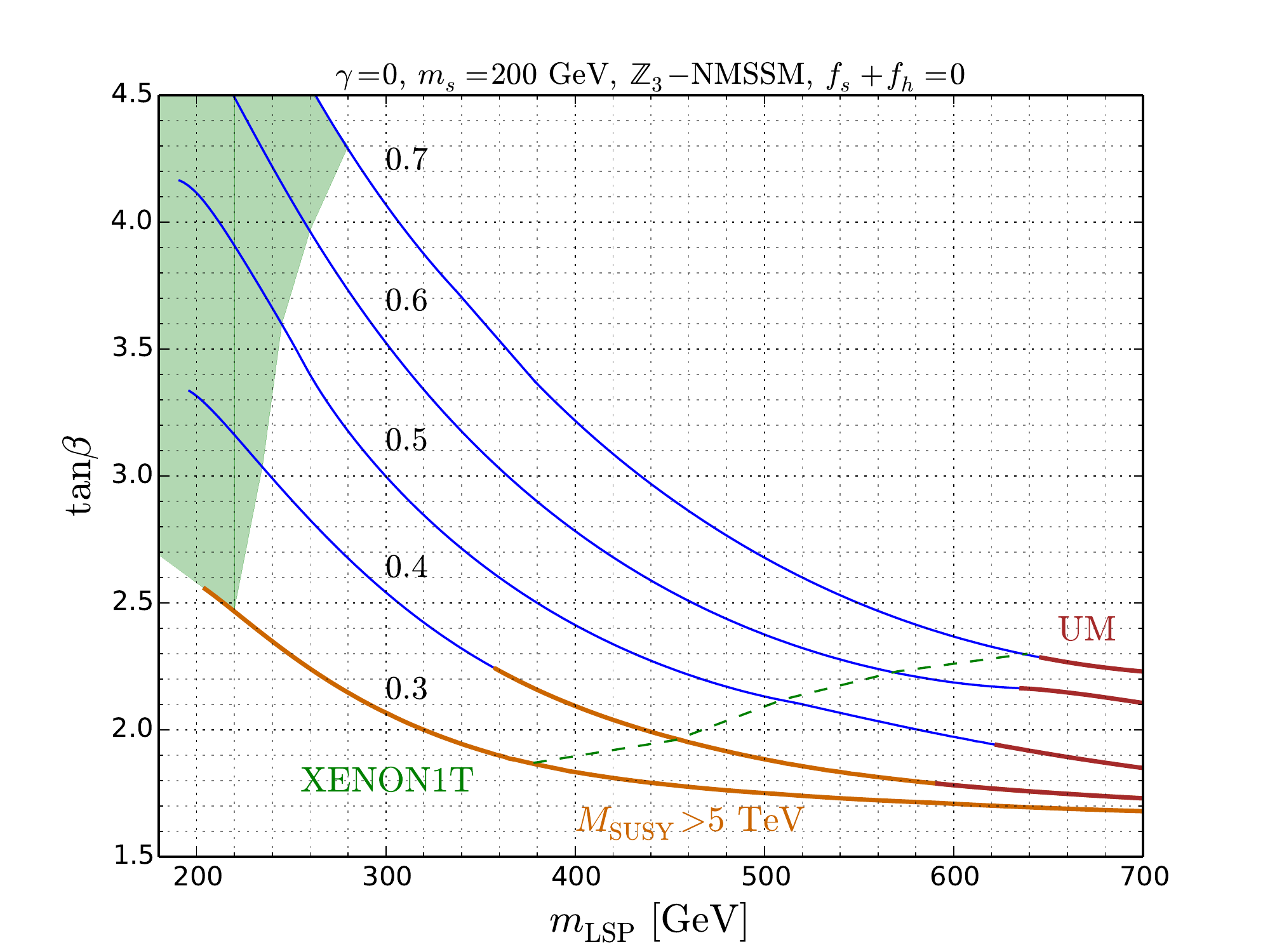}
\hspace{-5ex}
\includegraphics[width=0.49\textwidth]{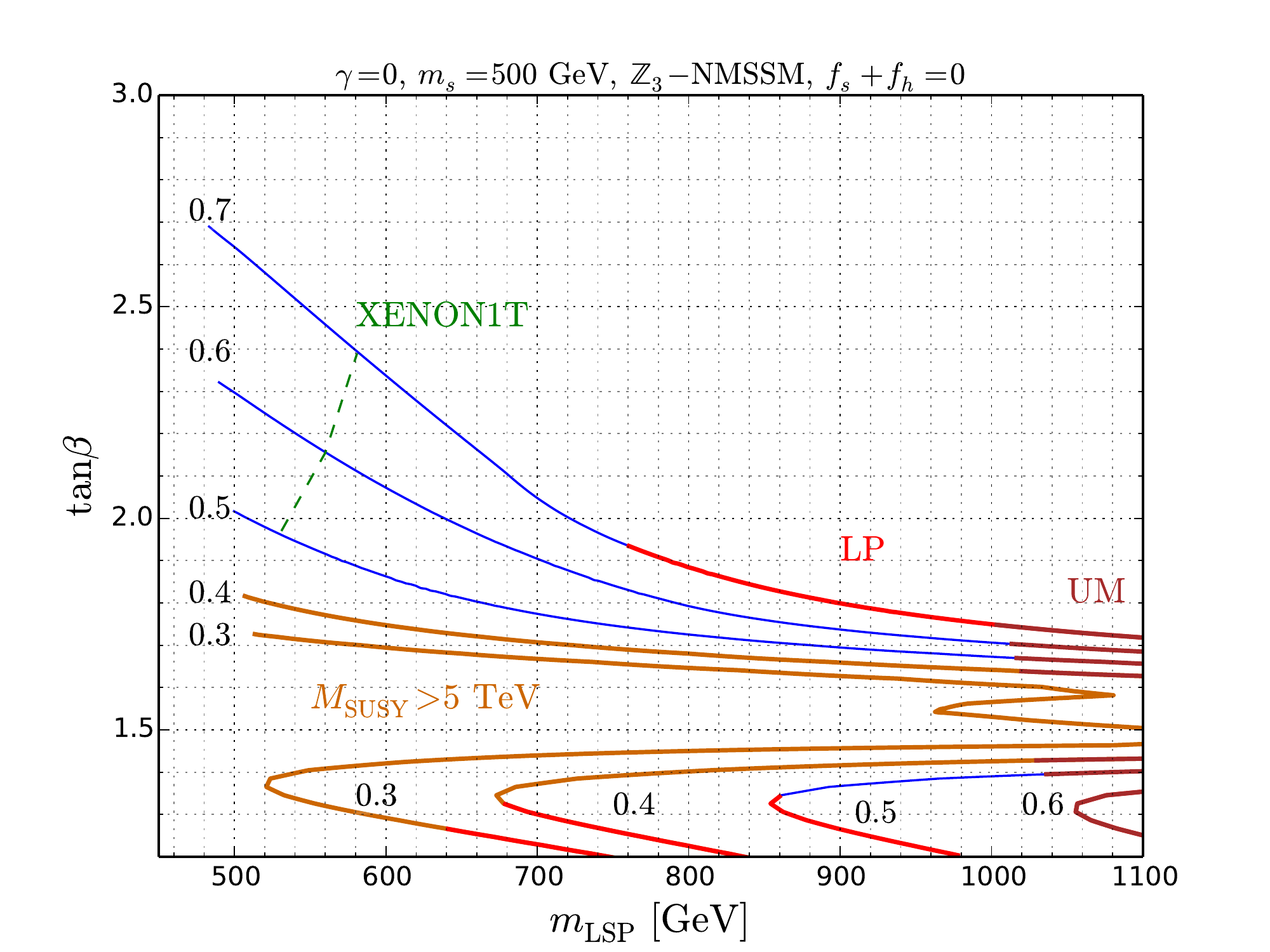}
\caption{Contours of $\Omega h^2=0.12$ in the plane $m_{\rm LSP}$-$\tan\beta$ assuming SI blind spot with $m_\chi>0$ and $\mu>0$ in the
$\mathbb{Z}_3$-invariant NMSSM for several values of $\lambda$ with $m_s=200$ (left panel) and $500$ GeV (right panel). The color code is the same as
in the right panel of Fig.~\ref{fig:channel_Z_om_0}. The lines denoted 
by UM depict regions in which {\tt NMSSMTools} reports unphysical global minimum.}
\label{Z3_heavy_s1}
\end{figure}

Let us first discuss the case of heavy singlet scalar in which only 
the Higgs exchange is relevant in the SI scattering amplitude and the 
SI blind spot has the standard form \eqref{bs_h_0}. In this case $|\gamma|$ 
must be close to zero to avoid large negative correction to the Higgs 
mass and eq.~\eqref{Z3cond2} implies $m_{\rm LSP}>m_s$. This is demonstrated 
in Fig.~\ref{Z3_heavy_s1} where it is clearly seen that for 
$m_{\rm LSP}\lesssim m_s$ there are no solutions (due to a tachyonic 
pseudoscalar).\footnote{In Fig.~\ref{Z3_heavy_s1} vanishing 
$\hat{h}$-$\hat{s}$ mixing is assumed but  for small non-zero mixing $\gamma$ (preferred by the Higgs mass)  the results are similar.  }

We also note that eq.~\eqref{Z3cond2} implies that resonant LSP annihilation 
via singlet-like scalar or pseudoscalar is typically not possible in this 
case.\footnote{Due to loop corrections to eq.~\eqref{Z3cond2} one may find 
some small regions of resonant annihilation mediated by a singlet for $m_s$ 
not far above $m_h$ and large $\lambda$ close to the perturbativity bound. W
e discuss this effect in more detail in subsection~\ref{subsec:Z3light}
because it is more generic for $m_s<m_h$.}
On the other hand, eq.~\eqref{Z3cond2} implies that 
the LSP annihilation channel into $sa$ via $a$ exchange is 
almost always open (for small mixing $\gamma$ and $m_s>m_h$ 
this channel is kinematically forbidden only in a small 
region of the parameter space for which $m_a\approx3m_s$).
This allows for smaller annihilation rate into $t\bar{t}$, hence also for smaller higgsino component of the LSP
and larger $\tan\beta$. In consequence, larger LSP masses consistent with $\Omega h^2=0.12$ and perturbativity up to the GUT scale are possible than
in the case with both singlets decoupled 
(compare Fig.~\ref{Z3_heavy_s1} to Fig.~\ref{fig:channel_Z_om_0}). For the same reason large enough LSP masses are beyond the reach of XENON1T, as
seen from Fig.~\ref{Z3_heavy_s1}.

\subsection{Light singlet scalar}
\label{subsec:Z3light}

The situation significantly changes when singlet-like scalar is light, 
especially if the Higgs-singlet mixing is not small (which enhances 
the Higgs mass if $m_s<m_h$). This is because the blind spot condition 
changes to eq.~\eqref{bs_fhs}. Moreover, for light singlet the loop 
corrections to condition \eqref{Z3cond2} can no longer be neglected 
which under some circumstances allows for resonant LSP annihilation via
the s-channel exchange of $a$.

\begin{figure}[t]
\center
\includegraphics[width=0.49\textwidth]{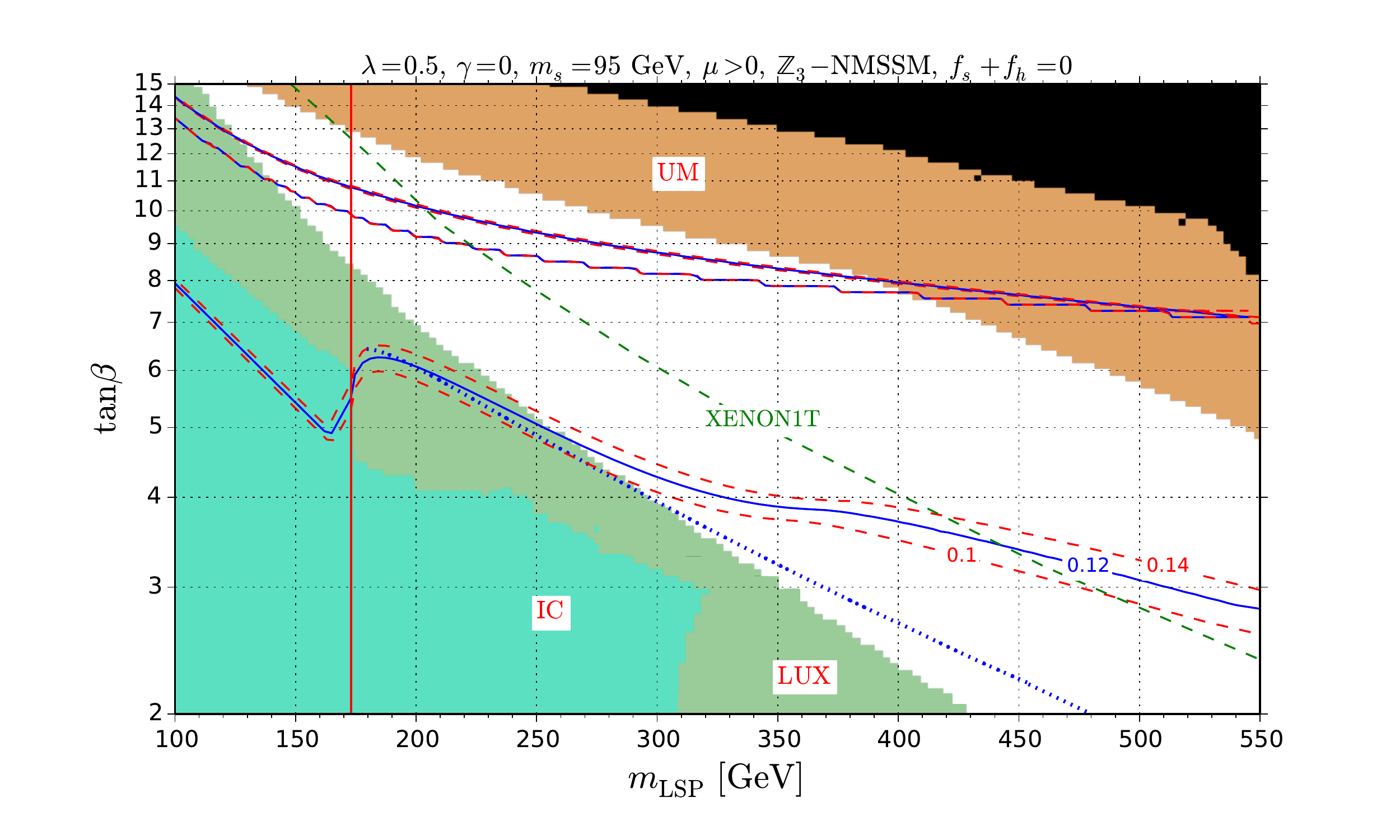}\hspace{-5ex}
\includegraphics[width=0.49\textwidth]{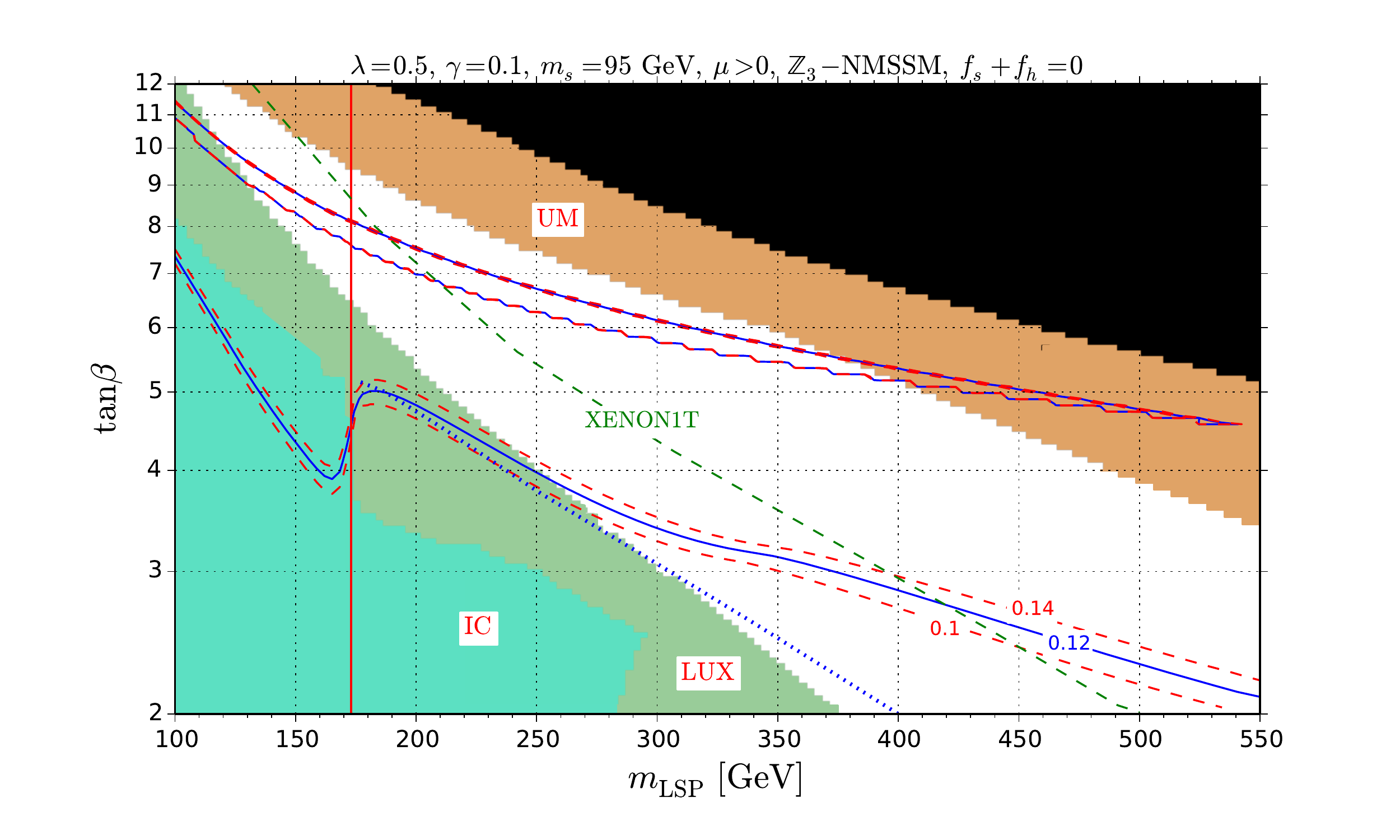}
\hspace{-5ex}
\includegraphics[width=0.49\textwidth]{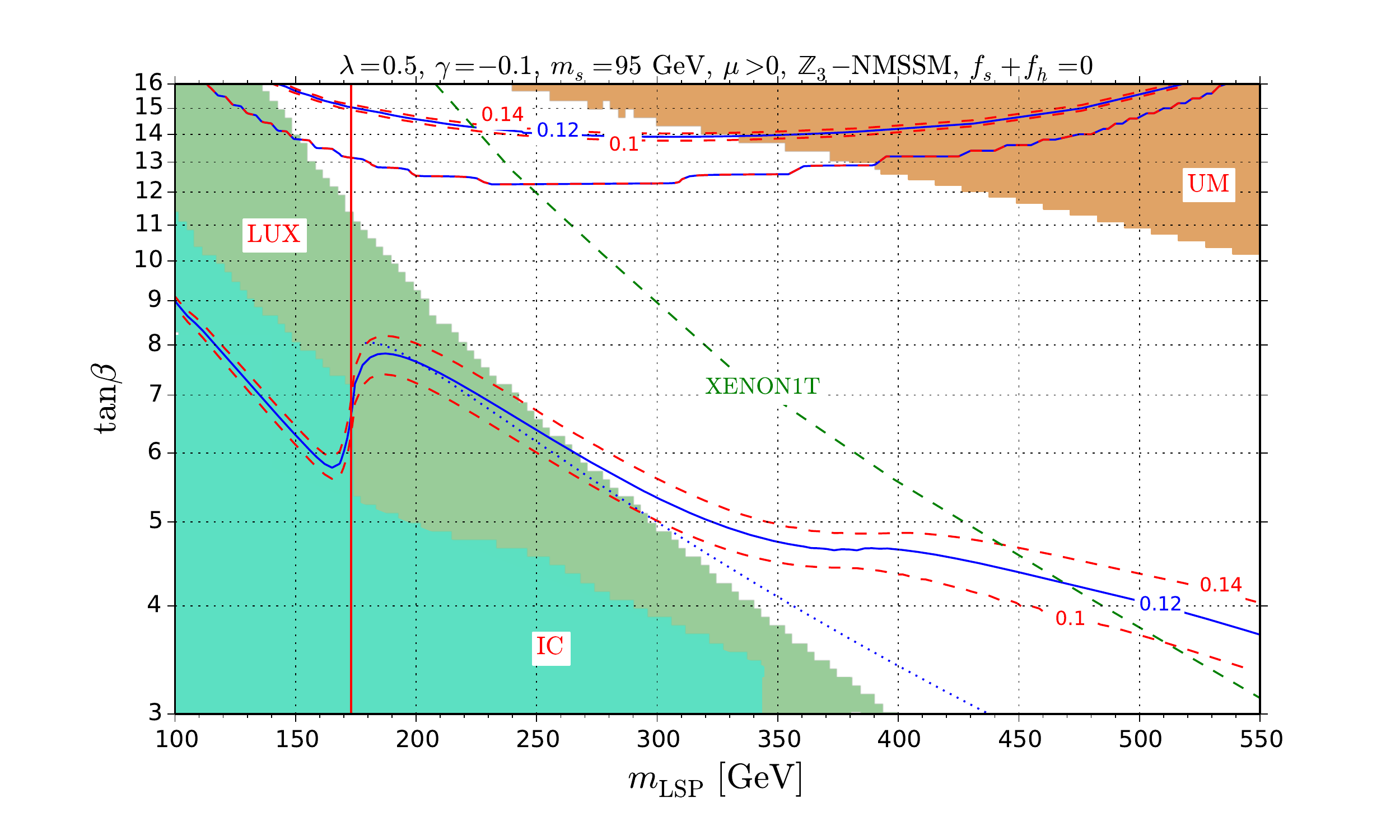}
\caption{The same as in Fig.~\ref{fig:cont_om_fhs} but in the $\mathbb{Z}_3$-invariant NMSSM and $\gamma=-0.1,0,0.1$.}
\label{fig:cont_om_Z3}
\end{figure}

\begin{figure}[t]
\center
\includegraphics[width=0.49\textwidth]{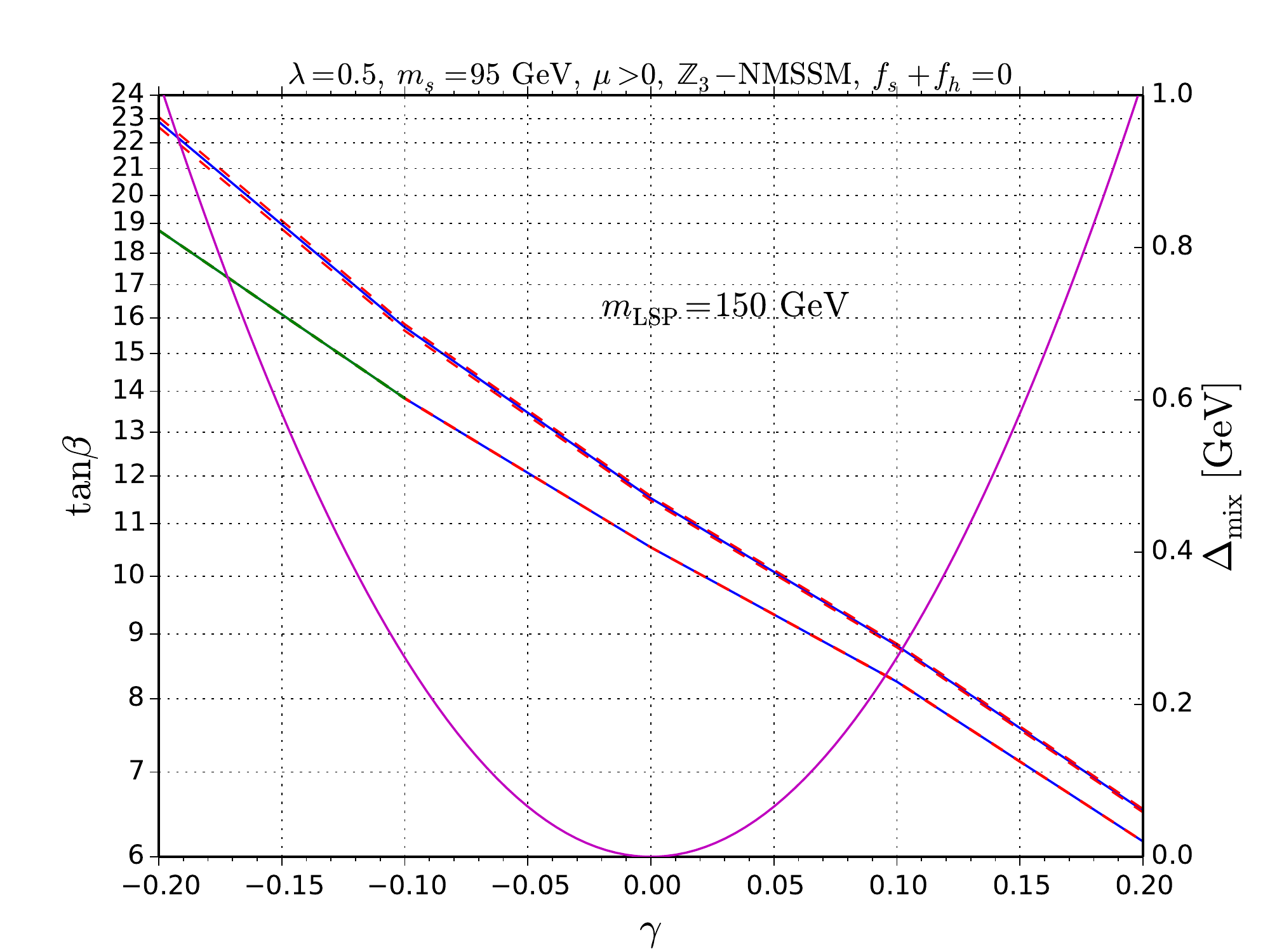}
\includegraphics[width=0.49\textwidth]{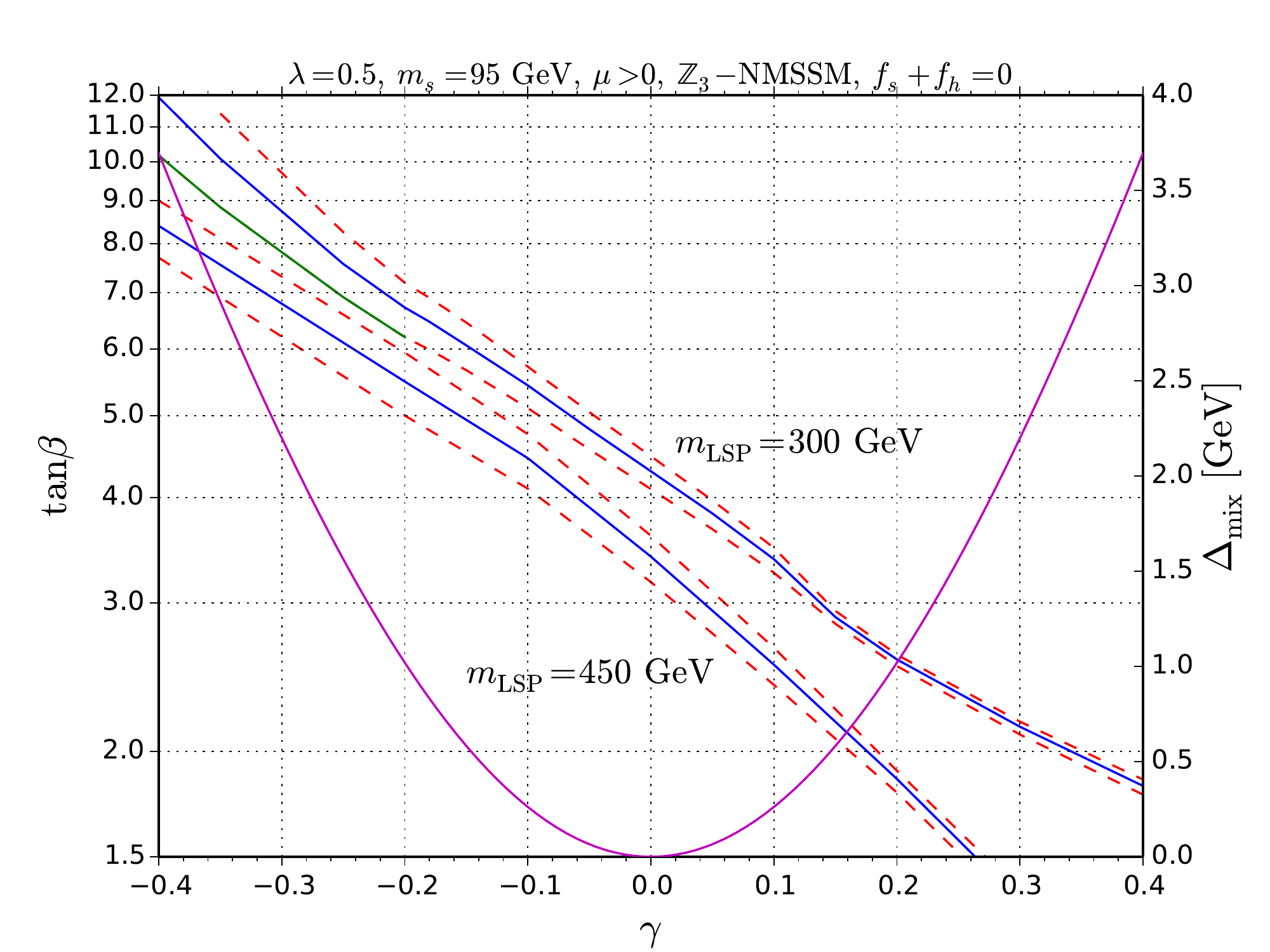}
\caption{Contour lines of $\Omega h^2=0.10,\,0.12,\,0.14$ as functions of $\tan\beta$ and $\gamma$ in the $\mathbb{Z}_3$-invariant NMSSM. Left panel: 
resonant annihilation via $a$ exchange for $m_{\rm LSP}=150$ GeV. Right panel: non-resonant annihilation for $m_{\rm
LSP}=300$ and $450$ GeV. The green parts of the contours are excluded by LUX. The parabola-like curves show dependence of $\Delta_{\rm mix}$ (on the
right horizontal axes) on $\gamma$. }
\label{fig:tanb_mLSP_Z3}
\end{figure}

In the $\mathbb Z_3$-symmetric NMSSM the singlet-dominated pseudoscalar $a$ 
plays quite important role for the relic density of the singlino-dominated 
LSP. First we check if and when the s-channel exchange of $a$ may 
dominate the LSP annihilation cross section and lead to the observed
relic density. Of course, this may happen if we are quite close to the 
resonance, i.e.~when $m_a\approx2\mlsp$. It occurs that it is not so easy 
to fulfill this requirement in the $\mathbb Z_3$-symmetric model. 
This is related to the condition \eqref{Z3cond2} which, for $m_a\approx2\mlsp$
and after taking into account the loop corrections in eqs.~\eqref{Mss} 
and \eqref{Maa}, may be rewritten in the form
\begin{equation}
\label{ms_Z3_res}
m_s^2 +\frac13 \mlsp^2 
+\gamma^2\left(m_h^2-m_s^2\right)
\approx
\Delta_{\hatss} + \frac13 \Delta_{\hataa}
\,.
\end{equation}
The l.h.s.~of the above expression is positive 
so this condition can not be fulfilled without the loop contributions. 
The last equation may be treated as a condition 
for the size of the loop corrections necessary to have resonant  
annihilation of the LSP mediated by the pseudoscalar $a$. 
In order to understand qualitatively the impact of condition 
\eqref{ms_Z3_res} on our analysis it is enough to consider the 
following simple situation: We assume that the scalar mixing $\gamma$
is negligible and the BS is approximated by \eqref{bs_h_0}.
On the r.h.s.~of eq.~\eqref{ms_Z3_res} we take into account only
the first term of the loop correction $\Delta_{\hatss}$ \cite{reviewEllwanger}
\begin{equation}
\label{Delta_ss}
\Delta_{\hatss}
\approx
\frac{1}{2\pi^2}\,\lambda^2 \mu^2 \ln\left(\frac{M_{\rm SUSY}^2}{\mu^2}\right)
+
\frac{1}{2\pi^2}\,\kappa^2 \mlsp^2 \ln\left(\frac{M_{\rm SUSY}^2}{\mlsp^2}\right)
\end{equation}
(the second term is subdominant because $\mlsp^2\ll\mu^2$ 
for a singlino-dominated LSP 
and $\kappa^2<\frac14\lambda^2$ due to condition \eqref{Z3cond1}).
In such approximation and for $\tan\beta\gg1$ condition \eqref{ms_Z3_res} 
simplifies to
\begin{equation}
\label{resonance_approx}
m_s^2 
\approx
\mlsp^2 \left[
\left(\frac{\lambda\tan\beta}{2\pi}\right)^2
\ln\left(\frac{2M_{\rm SUSY}}{\mlsp\tan\beta}\right) -\frac13
\right]
.
\end{equation}
For given values of $\lambda$ and $m_s$ any change of $\mlsp$ must be 
compensated by appropriate change of $\tan\beta$. 
The expression in the square bracket has a maximum as a function of $\tan\beta$ approximately at $1.2M_{\rm SUSY}/\mlsp$.
Thus, to keep the 
r.h.s.\ constant in order to stay close to the resonance one has to 
decrease $\tan\beta$ for small $\mlsp$ and increase for large $\mlsp$. In our numerical examples presented in Fig.~\ref{fig:cont_om_Z3} we fix $M_{\rm
SUSY}=4$~TeV so the maximum of the square bracket corresponds to $\tan\beta$ about 30 (10) for the LSP masses of 150 (500) GeV. For small $|\gamma|$
the $a$ resonance occurs at the blind spot for $\tan\beta$ of 
order 10. That is why $\tan\beta$ typically decreases with $\mlsp$, as can be seen from Fig.~\ref{fig:cont_om_Z3}. Local minimum for $\tan\beta$ is
present only in the lower panel of Fig.~\ref{fig:cont_om_Z3} because more negative values of $\gamma$ lead in general to larger $\tan\beta$ (see
Fig.~\ref{fig:tanb_mLSP_Z3} and discussion at the end of this section).
Nevertheless, in every case the $\Omega h^2=0.12$ curves corresponding to the $a$ resonance 
have horizontal-like behavior: do not change very much with 
the LSP mass (and have values of $\tan\beta$ of 
order 10 for $M_{SUSY}=4$ TeV that we use in our numerical examples).
This should be compared to the general case when 
such curves are almost vertical (narrow ranges of the LSP mass but 
wide ranges of $\tan\beta$) -- see Fig.~\ref{fig:cont_om_fhs}.
This difference comes from the fact that in the general model there 
are more parameters and eq.~\eqref{Z3cond2} is not fulfilled.

Fig.~\ref{fig:cont_om_Z3} shows that there are two situations for which
BS and correct value of DM relic density are still compatible with the 
latest bound on DM SD cross-section. One is the above discussed 
case of resonant annihilation with the light pseudoscalar exchanged 
in the s channel. The second one occurs for smaller $\tan\beta$ but 
bigger $m_{\rm LSP}$ and corresponds to annihilation via
non-resonant exchange of particles in the s channel. Usually the main 
contribution to the annihilation cross-section in such a case comes 
from the exchange of $Z^0$ boson decaying into $t\bar{t}$ final state. This process 
allows to avoid the LUX bounds on $\sigma_{\rm SD}$ for $m_{\rm LSP}
$ above about 300 GeV but is not sufficient to push $\sigma_{\rm SD}$ below
sensitivity of XENON1T, as discussed in section~\ref{sec:bs_fh}.
The situation changes when new final state channels, especially $as$, open.
Then not only the present bounds on $\sigma_{\rm SD}$  may be easily 
fulfilled but some parts of the parameter space are beyond the XEXON1T 
reach. We see from Fig.~\ref{fig:cont_om_Z3} that for light singlets the lower limit on the LSP mass from LUX may be relaxed to about 250 GeV. The
effect of annihilation into light singlets is even more important for even heavier LSP so XENON1T may not be sensitive to LSP masses above about 400
GeV.

In both cases discussed above the allowed values of $\tan\beta$ are 
correlated with the LSP mass. The exact form of such correlation depends 
on the $\hat{s}$-$\hat{h}$ mixing parameter $\gamma$. Quite generally values of 
$\tan\beta$ decrease with $\gamma$. 
This is illustrated in Fig.~\ref{fig:tanb_mLSP_Z3}
where the bands of allowed $\tan\beta$ as functions of $\gamma$ are
shown for a few values of the LSP mass. 
This correlation between $\tan\beta$ and $\gamma$ can be easily 
understood from eqs.~\eqref{eq:As}-\eqref{bs_fhs}. The first factor 
on the r.h.s~of \eqref{bs_fhs} grows in the first approximation
like $-\gamma$. This can not be compensated by decreasing $\kappa$ 
because in the $\mathbb{Z}_3$-symmetric NMSSM $\kappa$ is fixed 
by the LSP mass. The BS condition \eqref{bs_fhs} with increasing r.h.s.\
may be fulfilled by decreasing the absolute value of the negative
contribution to its l.h.s.\ i.e.\ by increasing $\tan\beta$.

%
%

\section{Conclusions}
\label{sec:concl}

Motivated by the recent strong LUX constraints we investigated consequences of the assumption that the spin-independent cross-section of
singlino-higgsino LSP scattering off nuclei is below the irreducible neutrino background. We determined constraints on the NMSSM parameter space
assuming that the LSP is a thermal relic with the abundance consistent with Planck observations and studied how present and future constraints on
spin-dependent scattering cross-section may probe blind spots in spin-independent direct detection.

In the case when all scalars except for the 125 GeV Higgs boson are heavy the new LUX constraints exclude the singlino-higgsino masses below
about 300
GeV
unless the LSP mass is very close to the half of the $Z^0$ boson mass (between about 41 and 46 GeV). In the allowed region LSP
dominantly annihilates to $t\bar{t}$ and $\tan\beta$ must be below about 3.5 (assuming perturbative values of $\lambda$ up to the GUT scale) with the
upper bound being stronger for smaller $\lambda$ and heavier LSP. There is
also an upper bound of about 700 GeV assuming perturbativity up to the GUT scale. We found that XENON1T has sensitivity to exclude the entire region
of dark matter annihilating dominantly to $t\bar{t}$. This conclusion apply to general models of singlet-doublet dark matter. On the other hand, the
LSP resonantly annihilating via $Z^0$ boson exchange is possible only for large $\tan\beta$ unless $\lambda$ is very small e.g. for $\lambda>0.5$,
$\tan\beta\gtrsim20$. Only small range of LSP masses around the resonance of about 2 GeV is beyond the XENON1T reach while LZ is expected to probe
$Z^0$ resonance completely.
In all of the above cases the LSP is dominated by singlino. Current and future constraints can be avoided also for very pure higgsino with mass in
the vicinity of 1.1 TeV. 

The situation significantly changes when singlet-like (pseudo)scalars are light. Firstly, the presence of light CP-even singlet scalar modifies the
condition for spin-independent blind spot when its mixing with other Higgs bosons is non-negligible. Depending on the sign of the mixing angle
between the singlet and the 125 GeV Higgs preferred values of $\tan\beta$ may be either increased or decreased, as compared to the case with heavy
singlet. Interestingly, $\tan\beta$ is increased when the Higgs coupling to bottom quarks is smaller than that to gauge bosons which is somewhat
favored by the LHC Higgs coupling measurements.

Secondly, the presence of light singlets opens new annihilation channels for the LSP. As a result, correct relic abundance requires smaller higgsino
component of the LSP which relaxes spin-dependent constraints. We found that  resonant annihilation via
exchange of singlet pseudoscalar is possible even in the $\mathbb{Z}_3$-invariant NMSSM. Interestingly, even far away from the resonant region the
lower limit on the mass of LSP annihilating mainly to $t\bar{t}$ may be relaxed to 250 GeV. For larger LSP masses $sa$ may become dominant
annihilation channel and the LSP masses above 400 GeV may be beyond the reach of XENON1T.

\section*{Acknowledgements}

This work has been partially supported by National Science Centre,
Poland, under research grants DEC-2014/15/B/ST2/02157,
DEC-2015/18/M/ST2/00054 and DEC-2012/04/A/ST2/00099, 
by the Office of High Energy Physics of the U.S. Department of Energy
under Contract DE-AC02-05CH11231, and by the National Science Foundation
under grant PHY-1316783. MB acknowledges support from the Polish 
Ministry of Science and Higher Education through its programme Mobility Plus (decision no.\ 1266/MOB/IV/2015/0).
PS acknowledges support from National Science Centre, Poland, 
grant DEC-2015/19/N/ST2/01697.

\appendix

\section{LSP-nucleon cross sections}
\label{App:cross-sections}

In this Appendix we collect several expressions useful 
in discussing the SI and SD cross-sections of LSP on nuclei.

The couplings of the $i$-th scalar to the LSP and to the 
nucleon, appearing in the formula \eqref{eq:sigSD} for 
the SI cross-section, 
after decoupling the gauginos are approximated, respectively, by
\begin{align}
\alpha_{h_i\chi\chi}
\approx
\sqrt{2}\lambda 
&\left[
\tilde{S}_{{h_i}\hat{h}}N_{15}\left(N_{13}\sin\beta+N_{14}\cos\beta\right)
+\tilde{S}_{{h_i}\hat{H}}N_{15}\left(N_{14}\sin\beta-N_{13}\cos\beta\right)
\right.\nn&\,\,\,\,\left.
+\tilde{S}_{{h_i}\hat{s}}
\left({N_{13}}{N_{14}}-\frac{\kappa}{\lambda}N_{15}^2\right)
\right]
\,,
\label{alpha-h00_G}
\end{align}
\begin{equation}
\label{alpha-hNN_G}
\alpha_{h_iNN}
\approx
\frac{m_N}{\sqrt{2}v} 
\left[
\tilde{S}_{{h_i}\hat{h}}\left(F^{(N)}_d+F^{(N)}_u\right)
+\tilde{S}_{{h_i}\hat{H}} \left(\tan\beta F^{(N)}_d-\frac{1}{\tan\beta} F^{(N)}_u \right)
\right]\,.
\end{equation}
The LSP couplings to pseudoscalars, important for the relic abundance 
calculation, are approximated by cite{reviewEllwanger}
\begin{equation}
\label{alpha-a00_G}
\alpha_{a_i\chi\chi}
\approx i\sqrt{2}\lambda\left[
N_{15}\tilde{P}_{i1}(\sin\beta\,N_{14}+\cos\beta\,N_{13})
+\tilde{P}_{i2}\left({N_{13}}{N_{14}}-\frac{\kappa}{\lambda}N_{15}^2\right)
\right]
,
\end{equation}
where $\tilde{P}_{ij}$ are elements of the matrix diagonalizing 
the pseudoscalar mass matrix defined in eq.~\eqref{tildeP}.

Parameter $\eta$ defined by eq.~\eqref{eta_def} and convenient 
for the discussion of SI blind, using eqs.~\eqref{Nj3Nj5} and \eqref{Nj4Nj5}, 
may be written in the form
\begin{equation}
\label{eta_lambda/mu}
\eta
=
\frac{\frac{\lambda v}{\mu}\,
\left(1-\left(\frac{m_{\chi}}{\mu}\right)^2\right)
\left(\frac{m_{\chi}}{\mu}-\sin2\beta\right)}
{
\left(\frac{\lambda v}{\mu}\right)^2
\left[\left(1+\left(\frac{m_{\chi}}{\mu}\right)^2\right)\frac{\sin2\beta}{2}-\frac{m_{\chi}}{\mu}\right]
-\frac{\kappa}{\lambda}\left(1-\left(\frac{m_{\chi}}{\mu}\right)^2\right)^2
}\,.
\end{equation}

With the help of eqs.~\eqref{Nj3Nj5} and \eqref{Nj4Nj5}, 
the combination of the LSP components crucial for $\sigma_{\rm SD}$,
$(N_{13}^2-N_{14}^2)$, may be written as:  
\begin{equation}
\label{eq:N13N14diff}
N_{13}^2-N_{14}^2=\frac{\left[1-\left(m_\chi/\mu\right)^2\right](1-N_{15}^2)\cos2\beta}{1+\left(m_\chi/\mu\right)^2-2\left(m_\chi/\mu\right)\sin2\beta}\,.
\end{equation}
We can see immediately that the cross-section disappears in the limit 
of $\tan\beta=1$ or a pure singlino/higgsino LSP. 
The ratio of the higgsino to the singlino components of the LSP
may be calculating from eqs.~\eqref{Nj3Nj5} and \eqref{Nj4Nj5}:
\begin{equation}
\label{Higgsino/singlino}
\frac{1-N_{15}^2}{N_{15}^2}
=
\left(\frac{\lambda v_h}{\mu}\right)^{\!\!2}
\frac{1+\left(m_\chi/\mu\right)^2-2{(m_\chi}/{\mu})\sin2\beta}
{\left[1-\left({m_\chi}/{\mu}\right)^2\right]^2}
\,.
\end{equation}
Using this relation we may rewrite formula \eqref{eq:N13N14diff} in the form:
\begin{equation}
\label{eq:N13N14diff_lambda}
N_{13}^2-N_{14}^2=\frac{\left[1-\left(m_\chi/\mu\right)^2\right]
\cos2\beta}{1+\left(m_\chi/\mu\right)^2-2\left(m_\chi/\mu\right)\sin2\beta
+\left[1-\left(m_\chi/\mu\right)^2\right]^2
\left({\mu}/{\lambda v}\right)^2}\,.
\end{equation}
%

\section{The LSP (co)annihilation channels}
\label{App:annihilation}

In this Appendix we will use the following expansion of $\sigma v$ around $v=0$:
\begin{equation}\label{sigmaV_expansion}
\sigma v=a+bv^2+\mathcal{O}(v^4)\,.
\end{equation}
Then, the relic density may be written as~\cite{Griest}:
\begin{equation}
\label{om_ab}
\Omega h^2\approx
\frac{9.4\times10^{-12}\;{\rm GeV^{-2}}\;x_f}{a+3b/x_f}\,,
\end{equation}
where $x_f\approx 25$.

\subsection{Resonance with the $Z^0$ boson (unitary gauge) \label{App:res_Z}}

Let us consider the LSP annihilation into the SM fermions (except the $t$ quark\footnote{The effect from the $t$ quark appears for $m_{\chi}\sim m_t$ which is quite far from the resonance~-- we will discuss this case separately in the next paragraph.})
via $Z^0$ exchange in $s$ channel. The expansion coefficient $a$ and $b$ in eq.~\eqref{sigmaV_expansion} are equal to:
\begin{equation}
\label{eq:res_Z_a}
a=\frac{g^4}{32\pi}\frac{(N_{13}^2-N_{14}^2)^2}{(4m_{\chi}^2-m_{Z^0}^2)^2+m_{Z^0}^2\Gamma_{Z^0}^2}
\left(1-\frac{4m_\chi^2}{m_{Z^0}^2}\right)^2
\times 
\sum_Fc_Fm_F^2\sqrt{1-\frac{m_F^2}{m_{\chi}^2}}\,,
\end{equation}
\begin{equation}
\label{eq:res_Z_b}
b\approx b_0=\frac{g^4}{32\pi}\frac{(N_{13}^2-N_{14}^2)^2}{(4m_{\chi}^2-m_{Z^0}^2)^2+m_{Z^0}^2\Gamma_{Z^0}^2}\times 
\frac{2m_{\chi}^2}{3}\sum_F c_F\,(2\beta_F^2-2\beta_F+1)\,,
\end{equation}
where $g\equiv\sqrt{(g_1^2+g_2^2)/2}$, $c_F=1$ for leptons and 3 for quarks, whereas $\beta_F=2|q_F|\sin\theta_{\rm W}^2$. The 0 index in $b_0$ parameter means that we put fermion masses to 0 (which is a very good approximation for $m_\chi\sim m_{Z^0}/2$; of course $a_0=0$). The sum over the SM fermions (except the $t$ quark) in~\eqref{eq:res_Z_b} equal $\sim 14.6$. It is worth noting that $b_0\sim m_{\chi}^2$ and $a\sim m_F^2(1-4m_\chi^2/m^2_{Z^0})$ which means that $b\gg a\sim 0$ (in contrary to naive expectation). Moreover, the terms proportional to higher powers of $v^2$ in $\sigma v$  (for $m_\chi\gg m_F$) are suppressed with respect to $bv^2$ term in geometric way by $v^2/4$. Therefore we can approximate $\sigma v\approx b_0v^2$ and hence expressed the relic density in the form of eq.~\eqref{om_ab}. We will however improve slightly this approach (see Appendix~\ref{App:relic_density}) and write our formula in the following form:
\begin{equation}
\label{eq:res_Z_omega_app}
\Omega h^2\approx 0.1\left(\frac{0.3}{N_{13}^2-N_{14}^2}\right)^2
\frac{m_{Z^0}^2}{4m_{\chi}^2}
\left[\left(\frac{4m_{\chi}^2}{m_{Z^0}^2}-1 +\frac{\bar{v}^2}{4}\right)^2+
\frac{\Gamma_{Z^0}^2}{m_{Z^0}^2}\right]\,.
\end{equation}
where the term proportional to $\bar{v}^2\approx 0.5^2$ stems from the fact that the dark matter particles posses some thermal energy during the
freeze-out. Eq.~\eqref{eq:res_Z_omega_app} reproduces very well the results obtained from \texttt{MicrOMEGAs} far from the resonance (see eg.
Fig.~\ref{fig:cont_om_0}), however very close to the resonance, especially for $m_\chi\lesssim m_{Z^0}/2$, the difference may be sizable
(Fig.~\ref{fig:channel_Z_om_0}).

\subsection{Annihilation into $t\bar{t}$ via $ Z^0$}

In this case the dominant contribution also comes from $Z^0$ exchange in $s$ channel but in contrary to the previous paragraph $m_\chi\sim m_F(= m_t)$. Therefore the statement that $b\gg a$ is now longer true. It becomes clear when we write down the expression for $a$ and $b$ terms in the limit $m_\chi\gg m_{Z^0}$:
\begin{equation}
\label{eq:ann_tt_a}
a\approx
\frac{3g^4}{32\pi}(N_{13}^2-N_{14}^2)^2
\frac{m_t^2}{m_{Z^0}^4}
\sqrt{1-\frac{m_t^2}{m_{\chi}^2}}\,,
\end{equation}
\begin{equation}
\label{eq:ann_tt_b}
b\approx
\frac{3g^4}{32\pi}(N_{13}^2-N_{14}^2)^2
\frac{m_t^2}{m_{Z^0}^4}
\frac{1}{4}
\left(1-\frac{m_t^2}{2m_\chi^2}\right)
\frac{1}{\sqrt{1-\frac{m_t^2}{m_{\chi}^2}}}\,.
\end{equation}
One can see that for $m_\chi\sim m_t$ both terms are comparable whereas for larger $m_\chi$ we have $a/b\approx 4$ and eq.~\eqref{eq:ann_tt_a} suffices (as we would expect, the terms proportional to higher powers of $v^2$ are suppressed for $m_\chi\gg m_t,m_{Z^0}$ as $v^2/4$). Similarly to eq.~\eqref{eq:res_Z_omega_app} we can find the expression for $\Omega h^2$. Combining~\eqref{om_ab} with \eqref{eq:ann_tt_a} and \eqref{eq:ann_tt_b} we get:
\begin{equation}
\label{eq:ann_tt_omega_app}
\Omega h^2\approx 0.1\left(\frac{0.05}{N_{13}^2-N_{14}^2}\right)^2
\left[
\sqrt{1-\frac{m_t^2}{m_{\chi}^2}}+
\frac{3}{4}\frac{1}{x_f}
\left(1-\frac{m_t^2}{2m_{\chi}^2}\right)
\frac{1}{\sqrt{1-\frac{m_t^2}{m_{\chi}^2}}}
\right]^{-1}\,.
\end{equation}
The above equation works well for $m_\chi\gtrsim 175$ GeV (see Fig.~\ref{fig:cont_om_0}), however for $m_\chi\approx m_t$ we have to be more careful because the expansion in $v^2$ breaks down. One can see that for $m_\chi\gg m_t,m_{Z^0}$ the square bracket in~\eqref{eq:ann_tt_omega_app}  equals roughly 1 and $\Omega h^2$ depends on $|N_{13}^2-N_{14}^2|$ only. Similarly to the case of the resonance with $Z^0$, the crucial experimental bounds comes from SD direct detection (see right plot in Fig.~\ref{fig:res_Z_limits}).

It is worth pointing out that both $a$ and $b$ coefficients in~\eqref{eq:ann_tt_a} and \eqref{eq:ann_tt_b} come purely from $-p^\mu p^\nu/m_{Z^0}^2$ term in $Z^0$ propagator. It was noticed long time ago~\cite{ann_tt_1,ann_tt_2} that taking into account this term is also crucial for DM annihilation in galactic halos ($v^2\approx 10^{-6}$) for $m_\chi\sim m_{Z^0}/2$. This is because the $a$ coefficient in~\eqref{eq:res_Z_a} vanishes which causes large dip in the annihilation cross section.

\section{Improved formula for $\Omega h^2$ near a resonance}
\label{App:relic_density}

The method described below may be found in~\cite{DuchGrzadkowski}. Let us consider a general expression for $\sigma v$ for scalar dark matter (with mass $m$) annihilating via $s$ channel exchange of a particle with mass $M$ and total decay width $\Gamma$: 
\begin{equation}
\label{sigmav_scalar}
\sigma v = \frac{\alpha}{(s-M^2)^2+\Gamma^2M^2}\,.
\end{equation}
For simplicity we assume we assume $\alpha = \rm const$ which is generally not the case, however we are mainly focused on the effect on $\Omega h^2$ coming from the denominator in~\eqref{sigmav_scalar}. Using dimensionless quantities $\delta\equiv 4m^2/M^2-1$, $\gamma\equiv\Gamma/M$ and considering non-relativistic approximation $s=4m^2/(1-v^2/4)\approx 4m^2(1+v^2/4)$ we get:
\begin{equation}
\label{sigmav_scalar_app}
\sigma v = \frac{\alpha/M^4}{(\delta+v^2/4)^2+\gamma^2}\,.
\end{equation}
Let us now define $Y(x)\equiv\frac{n}{s}$, where $x=T/m$, and write
\begin{equation}
\frac{1}{Y(\infty)} - \frac{1}{Y(x_d)}=
m\,M_{\rm Pl}\,\frac{g_s}{\sqrt{g}}\sqrt{\frac{\pi}{45}}
\int_{x_d}^\infty \frac{\langle\sigma v\rangle}{x^2}\;dx\,.
\end{equation}
Parameter $x_d$ is defined as a moment in thermal evolution of DM when the term $1/Y(x_d)$ starts to be small and can be safely neglected. Dark matter relic abundance can be then calculated by double integration over $v$ and $x$:
\begin{align}
\Omega h^2 &= \frac{2.82\cdot 10^8}{\rm GeV}m\,Y(\infty)\\
&\approx
\frac{2.82\cdot 10^8}{\rm GeV}\frac{1}{M_{\rm Pl}}
\frac{\sqrt{g}}{g_s}\sqrt{\frac{45}{\pi}}
\left[
\frac{1}{2\sqrt{\pi}}
\int_0^\infty dv\,(\sigma v)v^2
\int_{x_d}^\infty dx\,\frac{e^{-v^2x/4}}{\sqrt{x}}
\right]^{-1}
\end{align}
Note that we changed the usual order of integration. We will now perform the simpler integral over $x$, obtaining:
\begin{equation}
\frac{1}{2\sqrt{\pi}}
\int_0^\infty dv\,(\sigma v)v^2
\int_{x_d}^\infty dx\,\frac{e^{-v^2x/4}}{\sqrt{x}}=
\int_0^\infty(\sigma v)v\,{\rm erfc}(v/2\sqrt{x_d})\;dv\,.
\end{equation}
Substituting here eq.~\eqref{sigmav_scalar_app} and ${\rm erfc}(v/2\sqrt{x_d})\approx 1-\sqrt{x_d/\pi}\,v+\dots$ we can easily find simple expressions for $\Omega h^2$ for some hierarchical values of $\delta$ and $\gamma$ e.g. $\delta\ll\gamma$ etc.

In the case of fermionic dark matter our expression~\ref{sigmav_scalar} generalizes to:
\begin{equation}
\label{sigmav_fermion}
\sigma v = \frac{\alpha\;f(s)}{(s-M^2)^2+\Gamma^2M^2}\,.
\end{equation}
Now we have to perform the following integral
\begin{equation}
\label{Y_fermion}
...\frac{\alpha}{M^4}
\int_0^\infty 
\frac{f(v)v\;{\rm erfc}(v/2\sqrt{x_d})}
{(\delta+v^2/4)^2+\gamma^2}\;dv\,.
\end{equation}
In order to proceed further we have to specify the formula for $f(s)$. In the case of LSP annihilation into fermions via $Z^0$ exchange the dominant contribution is $f(v)\sim v^2$~-- see Appendix~\ref{App:res_Z}. Analytical form of the above integral is very complicated even for such simple expression for $f(v)$. The numerator in eq.~\eqref{Y_fermion} has a maximum for some specific value of $v$. Therefore we will take the denominator in front of the integral, substituting $v\to\bar{v}$, where $\bar{v}$ is defined as a mean value of the numerator. Then we have:
\begin{equation}
\label{Y_fermion_app}
...\frac{\alpha}{M^4}
\frac{1}{(\delta+\bar{v}^2/4)^2+\gamma^2}
\int_0^\infty 
v^3\;{\rm erfc}(v/2\sqrt{x_d})
\;dv=
...\frac{\alpha}{M^4}
\frac{3/x_d^2}{(\delta+\bar{v}^2/4)^2+\gamma^2}
\,,
\end{equation}
\begin{equation}
\label{bar_v}
\bar{v}\equiv
\frac{\int_0^\infty v\,v^3\;{\rm erfc}(v/2\sqrt{x_d})\;dv}
{\int_0^\infty v^3\;{\rm erfc}(v/2\sqrt{x_d})\;dv}
=\frac{64}{15\sqrt{\pi}}\;x_d^{-1/2}\,.
\end{equation}
For $x_d=25$ we have $\bar{v}\approx 0.5$. The above method effectively includes the fact that the dark matter particles posses some thermal energy during their freeze-out. Other cases of $f(v)$ can be also easily analyzed and compared with numerical results.


\end{document}